\documentclass{aa}  

\usepackage{chemmacros}

\newcommand{\ha}{H$\alpha$} 
\newcommand{\hb}{H$\beta$}

\newcommand{\hydrogeni}{H\,{\sc i}}

\newcommand{\helium}{He\,{\sc i}}

\newcommand{\nitrogen}{[N\,{\sc ii}]}

\newcommand{\nitrogena}{[N\,{\sc i}]}
\newcommand{\oxygeniii}{[O\,{\sc iii}]}

\newcommand{\oxygeni}{[O\,{\sc i}]}

\newcommand{\sulfurt}{[S\,{\sc ii}]}

\newcommand{\ironii}{[Fe\,{\sc ii}]}
\newcommand{\iron}{Fe}
\newcommand{\ironiii}{[Fe\,{\sc iii}]}

\newcommand{\silicon}{[Si\,{\sc i}]}

\newcommand{\degree}{$^{\circ}$}
\def\vhel{\ifmmode{V_{{\rm HEL}}}\else{$V_{{\rm HEL}}$}\fi}
\def\vsys{\ifmmode{V_{\rm sys}}\else{$V_{\rm sys}$}\fi}
\def\kms{\ifmmode{~{\rm km\,s}^{-1}}\else{~km~s$^{-1}$}\fi}
\def\vlsr{\ifmmode{v_{\rm lsr}}\else{$v_{\rm lsr}$}\fi}

\usepackage{graphicx}
\usepackage{txfonts}

\usepackage[colorlinks=true, urlcolor=blue, linkcolor=blue, citecolor=blue]{hyperref} 
\usepackage{tikz}

\definecolor{lime}{HTML}{A6CE39}
\DeclareRobustCommand{\orcidicon}{
        \begin{tikzpicture}
        \draw[lime, fill=lime] (0,0) 
        circle [radius=0.16] 
        node[white] {{\fontfamily{qag}\selectfont \tiny ID}};
        \draw[white, fill=white] (-0.0625,0.095) 
        circle [radius=0.007];
        \end{tikzpicture}
        \hspace{-2mm}
}

\foreach \x in {A, ..., Z}{\expandafter\xdef\csname orcid\x\endcsname{\noexpand\href{https://orcid.org/\csname orcidauthor\x\endcsname}
                        {\noexpand\orcidicon}}
}

\begin{document} 

   \title{\ironii~1.644~$\mu$m~imaging  survey of planetary nebulae with low-ionisation structures}


   \author{S. Akras\inst{1}{\orcidA{}}, I. Aleman\inst{2}{\orcidB{}}, D. R. Gon\c{c}alves\inst{3},  G. Ramos-Larios\inst{4}{\orcidC{}}, K. Bouvis\inst{1}{\orcidD{}}}

    \authorrunning{S. Akras et al.}

   \institute{Institute for Astronomy, Astrophysics, Space Applications and Remote Sensing, National Observatory of Athens, GR 15236 Penteli, Greece,
            \email{stavrosakras@gmail.com}
         \and
             Instituto de F\'isica e Qu\'imica, Universidade Federal de Itajub\'a, Av. BPS 1303, Pinheirinho, 37500-903, Itajub\'a, MG, Brazil  
        \and
             Observat\'orio do Valongo, Universidade Federal do Rio de Janeiro, Ladeira Pedro Antonio 43, 20080-090, Rio de Janeiro, Brazil
        \and 
            Instituto de Astronom\'ia y Meteorolog\'ia, CUCEI, Av. Vallarta No. 2602, Col. Arcos Vallarta, CP 44130, Guadalajara, Jalisco, Mexico
             }
   \date{}

 
  \abstract
    {Low-ionisation structures (LISs) are commonly found in planetary nebulae (PNe), but they are still poorly understood. The recent discovery of unforeseen molecular hydrogen gas (H$_2$) has impacted what we think we know about these microstructures and PNe. To obtain an overall understanding of LISs, we carried out an \ironii~1.644~$\mu$m imagery survey in PNe with LISs, with the aim to detect the \ironii~1.644$\mu$m emission line, a common tracer of shocks. We present the first detection of \ironii~1.644~$\mu$m line directly associated with the LISs in four out of five PNe. The theoretical H I 12-4 recombination line was also computed either from the Br$\gamma$ or the H$\beta$ line and subtracted from the observed narrow-band line fluxes. The \ironii~1.644~$\mu$m flux ranges from 1 to 40 $\times$10$^{-15}$ ergs~cm$^{-2}$~s$^{-1}$ and the intensity from 2 to 90 $\times$10$^{-5}$~erg~s$^{-1}$~cm$^{-2}$~sr$^{-1}$. The R(\iron)=\ironii~1.644~$\mu$m/Br$\gamma$ line ratio was also computed and found to range between 0.5 and 7. In particular, the \ironii~1.644~$\mu$m line was detected in NGC~6543 (R(\iron)$<$0.15), along with the outer pairs of LISs in NGC~7009 (R(\iron)$<$0.25) and the jet-like LISs in IC~4634 (R(\iron)$\sim$1), and in several LISs in NGC~6571 (2$<$R(\iron)$<$7). The low R(\iron) result for NGC~6543 is  attributed to the UV radiation from the central star. In contrast, the higher values in NGC~6571 and IC~4634 are indicative of shocks. The moderate R(\iron) in NGC~7009 likely indicates the contribution of both mechanisms.}


\maketitle
%
%
\section{Introduction}

Planetary nebulae (PNe) are the result of the complex interaction between stellar winds (interacting stellar wind model-ISW, \citealt{kwok1978}, and generalised interacting stellar winds model-GISW, \citealt{kwok1982,balick1987}), the ionisation and excitation processes of the expelled gas illuminated by the intense far-ultraviolet (far-UV) radiation from the central stars and the supersonic shock waves that propagate outward into the AGB circumstellar medium \citep[e.g.][and references therein]{Mellema2004,Perinotto2004,Schonberner2005a,Schonberner2005b}.

The illumination of the circumstellar gas by highly energetic UV photons is the main heating mechanism in PNe, yet shocks can also act as an extra heating process. Electron temperature augmentation is corroborated by a noticeable enhancement of the \oxygeniii/\ha~line ratio and may indicate shock interaction \citep[e.g.][]{Guerrero2013}. Collimated fast outflows that are frequently found in PNe \citep[e.g.][]{Akras2012,Akras2015,Miranda2017,Sabin2017,Sowicka2017,Derlopa2019,Rechy2020} are typical exemplars of shock-heated structural components. Several studies have demonstrated that collimated outflows or jets are strongly related to the formation of aspherical PNe, toroidal and ring structures, and dense knots \citep[e.g.][among others]{Sahai1998,Akashi2008,sahai2011,akashi2017,akashi2018,Balick2018,Balick2020,Bermudez2020,Segura2020,Soker2020,Akashi2020}. A recent statistical analysis of collimated outflows and jets in PNe has shown that the bulk of their velocity is around 60~\kms\  and the host nebulae are typically young \citep[$<$3000~yrs;][]{Guerrero2020}. 

The enhancement of low-ionisation emission lines such as \nitrogen~$\lambda\lambda$6548,6584, \sulfurt~$\lambda\lambda$6716,6731, and \oxygeni~$\lambda\lambda$6300,6363 at distinct microstructures in PNe  (e.g. knots, filaments; hereafter low-ionisation structures, LISs, e.g.~\citealt{Goncalves2001,Belen2023b}) has been attributed to the intense UV radiation field from the central star \citep[e.g.][]{Ali2017} or in conjunction with a shock interaction \citep[e.g.][]{Dopita1997,Phillips1998,goncalves2004,Goncalves2009,Akras2016,Belen2023,Belen2023b}. In particular, the intensity of the \oxygeni~$\lambda$6300 and \sulfurt~$\lambda\lambda$6716,6731 emission lines are stronger than the photoionisation process predicts and they are attributed to a shock-heated gas \citep{Phillips1998}. Although the \sulfurt/\ha~line ratio has been widely used to identify supernova remnants (shock-dominated) and distinguish them from PNe and H~{\sc ii}~regions that are UV-dominated \citep[e.g. ][]{Leonidaki2013,Sabin2013,Fesen2015,Kopsacheili2020,Akras2020abell}, this ratio is not straightforwardly applicable to trace shocks in PNe \citep[e.g.][]{Belen2023b}. Numerical simulations have demonstrated that models of ionised gas with low photo-ionisation rates and shock-heated gas can generate very similar spectroscopic characteristics \citep{Raga2008} (see also \citealt{Kopsacheili2020}). This makes very hard to determine the contribution of shocks in PNe dominated by UV radiation through optical emission lines only.

In the infrared (IR) wavelength regime, H$_2$ 1-0~S(1)/2-1~S(1) $\geq$4 and H$_2$ 1-0~S(1)/Br$\gamma$ $\geq$10 line ratios are also considered as shock indicators for molecular gas \citep[e.g.][]{Beckwith1980,MarquezLugo2015}, while the lower values are associated with UV radiation \citep{Aleman2011a,Aleman2011b}. However, a high density medium ($\geq$10$^{4-5}$~cm$^{-3}$) illuminated by intense UV radiation field can imitate the line ratios of shock-heated gas, considering that the collisional de-excitation process becomes significant \citep{Sternberg1989}. Recently, \cite{Aleman2020} claimed that the large H$_2$ 1-0~S(1)/2-1~S(1) and H$_2$ 1-0~S(1)/Br$\gamma$ observed ratios from several PNe can be naturally explained by photoionisation process if the slit configuration used in the observations is taken into account. Overall, the detection of the H$_2$ 1-0~S(1) emission line directly originated from the LISs in some PNe \citep{Akras2017,Akras2020} did not provide a clear answer regarding the dominant mechanisms, as both UV radiation and collisions by shock waves are able to reproduce the observations.

Other common tracers of shocks in IR wavelengths are the \ironii~emission lines centered at 1.257 and 1.644~$\mu$m. They have been detected in a wide variety of sources such as active galactic nuclei (AGNs), supernova remnants, Herbig-Haro objects (HH), and PNe. In particular, in PNe, the \ironii~1.644~$\mu$m line has been mainly detected with spectroscopic observations \citep{Lumsden2001}. 

Shock waves very likely destroy the dust grains and liberate iron into the gas phase, resulting in stronger \ironii~lines compared to an unshocked photoionised gas \citep{Graham1987}. In particular, the R(\iron)=\ironii\ 1.644~$\mu$m/Br$\gamma$ line ratio $>$~0.1 has been widely considered as an indicator of shock activity. Yet, the \ironii\ 1.644~$\mu$m line can also be found in photoionised gas such as the Orion H~II region, where R(\iron)~$<$~0.06 \citep{Lowe1979}, and PNe \citep{Lumsden2001}. The comparable ionisation potentials of Fe\ch{^{+}} (16.2~eV) and O\ch{^{0}} (13.6~eV) implies co-spatially \ironii~1.644~$\mu$m and \oxygeni~6300\AA~emission lines, both emanate from the same partially ionised region. The \ironii/Br$\gamma$--\oxygeni/H$\alpha$ diagnostic diagram as well as the excitation mechanisms of the iron lines by UV radiation and collisions (shocks) on Seyfert and starburst galaxies have been discussed by \cite{Mouri1990,Mouri2000}. All the aforementioned optical and IR tracers of shocks, along with other potential tracers, are discussed in \cite{Hollenbach1989}.

With respect to planetary nebulae, it has not been reported any detection of iron emission lines directly associated with the LISs, despite their enhanced \oxygeni~6300\AA~emission due to the lack of spatially resolved images.
In this paper, we present a pilot narrow-band \ironii~1.644~$\mu$m imaging survey of PNe with particular interest in their LISs. The paper is organised as follows. In Sect.~2, we make a brief review of the \ironii~1.644~$\mu$m emission line detections in PNe. The sample selection and the observations are described in Sects.~3 and 4, respectively. The contribution of the \hydrogeni~12-4~emission line centered at 1.640~$\mu$m is also discussed. The results of the survey are discussed in Sect.~5 and we finish with our conclusions in Sect.~6.

\section{[Fe~{\sc ii}]~1.644~$\mu$m emission line in PNe}

Spectroscopic observations in the H and K bands for a number of PNe were presented  by \cite{Hora1999,Lumsden2001} and \cite{Likkel2006}. In some of the spectra, the \ironii~1.644~$\mu$m line is blended with the \hydrogeni~12-4~1.640~$\mu$m line due to the low spectral resolution of the spectrograph and its detection is questionable.

\begin{table}
\centering
\caption{Ages of planetary nebula where the \ironii~1.644~$\mu$m emission line was detected.}
\label{tab:Fe_detections}
\begin{tabular}{lcl}

\hline                                  
PN      &       Age       &     References      \\
name    &   (years)   &      \\
\hline                                  
CRL~618 &       $\sim$200       &       1 \\
M~2-9   &       1200-2000       &       2       \\
NGC~2392 (outer shell)  &       $\sim$9300      &       3       \\
NGC~2392 (jets) &       1800-2600       &       3,4\\
NGC~2440        &       $\sim$1000      &       5 \\
NGC~7027        &       700-2000        &       6,7 \\
Hb~12 (outer knots)     & 2860-3250     &       8\\
Hb~12 (hourglass nebula) & $\sim$1120 & 8\\
Hb~12 (compact central region) & $\sim$300 & 9\\
BD~+30\degree~3639      &       600-800 &       10,11,12\\
Vy~2-2  &       $\sim$200       &       13,14   \\
M~1-92  &       $\sim$900       &       15\\
M~4-18  &       $\sim$3100      &       16\\
M~1-16  &       $\leq$850       &       17\\
NGC~7009 (main shell) &  1650 &  18 \\
NGC~7009 (outer shell) & 2200 &  18 \\
NGC~7009 (outer pair of knots) & $\sim$900 &  19 \\
NGC~6543 (E25) &    $\sim$1040  & 20  \\
NGC~6751 (inner halo) & $\sim$12800 &   21\\
NGC~6751 (ring knots) & $\sim$5100  &   21 \\
NGC~6751 (bipolar lobes) & $\sim$2700 &         21 \\
NGC~6302 &  $\sim$2100 & 22,12\\          
IC~4634  &   $\sim$3210 & 23\\
\hline
\end{tabular}
\tablebib{
(1)~\citet{Bujarrabal1988}; (2) \citet{Smith2005}; (3) \citet{GarciaDiaz2012}; (4) \citet{Guerrero2021};
(5) \citet{Wolff2000}; (6) \citet{Latter2000}; (7) \citet{santander2012}; (8) \citet{Vaytet2009};
(9) \citet{Miranda1989}; (10) \citet{Li2002}; (11) \citet{akras2012BD};(12)~\citet{GomezGordillo2020}; (13) \citet{miranda1991}; (14) \citet{Christiano1998}; (15) \citet{Buharrabal1998}; (16) \citet{DeMarco1999}; (17) \citet{Sahai1994}; (18) \citet{Sabbadin2004}; (19) \citet{Fernandez2004}; (20) \citet{Reed1999}; (21) \citet{Clark2010}; (22) \citet{Wright2011}; (23) \citet{Guerrero2020}.
}
\end{table}

Nevertheless, the theoretical flux of the \hydrogeni~12-4 line can be derived from the Br$\gamma$ flux and then subtracted from the \ironii+\hydrogeni~12-4 blended observed flux. Considering the Case B recombination theory, an electron density (n$_{\rm e}$) of 10$^4$~cm$^{-3}$ and electron temperature (T$_{\rm e}$) of 10$^4$~K, the theoretical flux of \hydrogeni~12-4 line is approximately 19 percent of the Br$\gamma$ flux \citep[\hydrogeni~12-4~/~Br$\gamma$~$\sim$0.19,][]{Hummer1987}.


For the majority of the PNe in the sample of \cite{Lumsden2001}, we find that the blended \ironii+\hydrogeni~12-4 intensity is consistent with the expected flux of the \hydrogeni~12-4 line. Therefore, the \ironii~1.644~$\mu$m line is certainly detected via spectroscopy only in the following PNe: CRL~618, BD~+30\degree~3639, NGC~7027, M~4-18, M~1-16, M~1-92, M~2-9, Vy~2-2, and M~1-78. 

Besides the aforementioned spectroscopic data,  \ironii~narrow-band images have been obtained only for a handful of pre-PNe and PNe: Hubble~12, M~2-9, CRL~618, NGC~6302, and NGC~7027. Hubble~12 displays strong \ironii~1.644~$\mu$m emission from an inner hourglass structure with an intensity around 8$\times$10$^{-15}$ ergs cm$^{-2}$ s$^{-1}$ arcsec$^{-2}$, being engulfed by a more extended H$_2$ envelope \citep{Hora1996,welch1999,Clark2014}. The contribution of the \hydrogeni~12-4 emission line to the total emission was taken into consideration, and the \ironii\ emission was found to be consistent with J-shock models of $\sim$100~km~s$^{-1}$. The core of the nebula exhibits R(\iron)$<$0.1 as it is expected for a photodominated gas, while the spectroscopic data obtained with an offset from the core result in higher ratios of 0.14 \citep{Luhman1996} and up to 3 \citep{Hora1996}, implying a shock-heated gas.

The intriguing PN M~2-9 has also been imaged in the \ironii~1.644~$\mu$m line and a collimated bipolar structure enclosed by an alike but larger H$_2$ and ionised structure has been unveiled \citep[][]{Smith2005, Clyne2015, Balick2018}. Interestingly, the young PN K~4-47 displays a very similar H$_2$ structure \citep{Akras2017} but no \ironii\ images are available yet. According to the spectroscopic data of M~2-9, \ironii~1.644~$\mu$m emission is stronger in the northern knot and bipolar lobes in comparison to the core. The assumed mechanism responsible for the emission is shocks \citep{Hora1994}.

The highly collimated pre-PN CRL~618 is the third nebula imaged in the \ironii~1.644~$\mu$m line, displaying emission at the edges of the outflows. Because of the highly expanding outflows with velocities of 300~km s$^{-1}$, the \ironii~1.644~$\mu$m emission has also been attributed to shock-heated gas  \citep{Balick2013}. Shocks have also been proposed as the principal mechanism responsible for the excitation of H$_2$ emission along the outflows of CRL~618 \citep[][]{Cox2003,Kastner2001}. Unfortunately, the low spatial resolution of the H$_2$~image does not allow for any comparison with the higher spatial resolution \ironii~image to be performed. 

The \lq Butterfly nebula\rq~NGC~6302 \citep{Kastner2022} and the \lq Jewel Bug nebula\rq~NGC~7027 \citep{Moraga2023} have recently mapped in the \ironii~1.644~$\mu$m emission line using the HST and the narrow-band filter F164N. The \ironii~1.644~$\mu$m line is clearly detected at the edges of the bipolar outflows (\lq wedge\rq) in NGC~6302, displaying an S-shape structure with a peak intensity of $\sim$3$\times$10$^{-14}$ erg~cm$^{-2}$~s$^{-1}$~sr$^{-1}$ \citep{Kastner2022}. The same authors have argued that the dominant excitation mechanism is shocks between a fast stellar wind with a velocity of $>$100~km~s$^{-1}$ and the previously formed bipolar lobes. HST \ironii~1.644~$\mu$m narrow-band images of NGC~7027 have also been presented by \cite{Moraga2023}. The \ironii~1.644~$\mu$m~emission is detected at the tips of a symmetric outflow along the south-east and north-west direction. The Chandra observations of NGC~7027 have revealed the presence of a very hot bubble with temperature of up to 3.6~MK. The X-ray emission detected in this nebula is also extended along the same direction \citep{Rofolfo2018}. A shock interaction between the collimated outflow and the nebular shell have been proposed to be responsible for this \ironii~emission \citep{Moraga2023}.

Overall, the \ironii~1.644~$\mu$m emission line is undoubtedly found only in sixteen PNe or pre-PNe, morphologically characterised as bipolar with collimated outflows or jet structures. It is worth noticing that most of these PNe are younger than 3000-3500~years, while the \ironii~1.644~$\mu$m line is more commonly detected in nebular components younger than 1000-1500~years (see Table~\ref{tab:Fe_detections}).


\section{Sample selection}

Five PNe with LISs were selected for this pilot narrow-band \ironii~1.644~$\mu$m imaging survey. All but one PN (NGC~6751) are described as elliptical with collimated outflows. NGC~6751 is a complex multi-shell PNe with an equatorial ring fragmented in knots and a pair of faint jet-like features \citep{Clark2010}.

Based on HST \oxygeniii/\ha~line ratio maps, \cite{Guerrero2013} classified four of them (NGC~6201, NGC~7009, NGC~6543, and IC~4634) as types A/B. The common characteristic of these types is the enhancement of \oxygeniii~emission line at the tips of collimated outflows \citep[see Fig.1 in][]{Guerrero2013}. NGC~6751 was not included in the previous analysis, but we assigned a D-type\footnote{\cite{Guerrero2013} define the D-type PNe as those with a nearly flat \oxygeniii/\ha~ratio map and a complex structure that does not fit with the other types.} classification because of its similarity with other PNe from the same type (e.g. NGC~2392 or NGC~6369).

Two of the PNe (NGC~7009 and NGC~6543) in our sample display H$_2$ emission that emanate from their LISs \citep{Fang2018,Akras2020}. However, the H$_2$ 1-0/H$_2$ 2-1 and H$_2$ 1-0/Br$\gamma$ ratios do not allow for  the origin of the H$_2$ emission (UV-pumping or collisional excitation) to be traced. No information about any molecular gas in the remaining three PNe is available yet.

\section{Observations}

A near-IR (NIR) narrow-band imaging survey of PNe with LISs was carried out in 2017 (February 20, March 25, June 6 and 9, and July 2 and 3, Program ID: GN--2017A--Q--58, PI: S. Akras) in service mode, using the Near InfraRed Imager and Spectrometer on the Gemini-North Telescope at Mauna Kea in Hawaii.

 
The narrow-band filter G0215 was used to isolate the \ironii\ line centered at 1.644~$\mu$m, while the narrow-band H-continuum filter (G0214) at 1.570~$\mu$m was used to determine and finally remove any adjacent continuum emission. The contribution of the \silicon\ line centered at $\lambda$1.645~$\mu$m to the total emission is expected to be negligible or even totally absent \citep[e.g.][]{Lumsden2001}. On the other hand, the contribution of the \hydrogeni~12-4 line ($\lambda$1.640~$\mu$m) to the total emission may be significant, and it is necessary to subtract it. Hereafter, we refer to the resultant observed continuum-subtracted images of our survey as \ironii+\hydrogeni~12–4~images, and the \hydrogeni~12-4 subtracted one as \ironii~1.644~$\mu$m images.

\begin{table}
\caption{Observations log} 
\centering
\label{table1}
\begin{tabular}{lcccc}
\hline
\multicolumn{1}{c}{PN} & \multicolumn{2}{c}{\ironii~1.644~$\mu$m +} & \multicolumn{2}{c}{H-continuum$^{\dag\dag}$}\\
& \multicolumn{2}{c}{\hydrogeni~12-4~1.640~$\mu$m$^{\dag}$} & \\
\hline
\multicolumn{1}{c}{} & \multicolumn{1}{c}{T$^{a}$} & \multicolumn{1}{c}{$\#$} & \multicolumn{1}{c}{T$^{a}$} & \multicolumn{1}{c}{$\#$} \\
\multicolumn{1}{c}{ } & \multicolumn{1}{c}{(s)} & \multicolumn{1}{c}{Frames} & \multicolumn{1}{c}{(s)} & \multicolumn{1}{c}{Frames} \\
\hline
NGC~6210  &  177  & 10  &  177 & 9 \\
NGC~6751  &  176$^b$  & 11 & 176 & 10 \\
NGC~6543  &  180$^c$  & 8 & 180$^c$  & 10 \\ 
NGC~7009  &  177 & 10 &  177 & 10 \\
IC~4634   &  170 & 7 &  170 & 7 \\
\hline
\end{tabular}
\tablefoot{
\tablefoottext{\dag}{$\rm{\lambda_c}$~=~1.644~$\mu$m, $\rm{\Delta\lambda}$~=~0.02466~$\mu$m.}
\tablefoottext{\dag\dag}{$\rm{\lambda_c}$~=~1.570~$\mu$m, $\rm{\Delta\lambda}$~=~0.02355~$\mu$m.}
\tablefoottext{a}{Exposure time of each frame in s.}
\tablefoottext{b}{2 coadds of 88~s each.}
\tablefoottext{c}{9 coadds of 20~s each.}
}
\end{table}

\begin{figure*}
\includegraphics[scale=0.275]{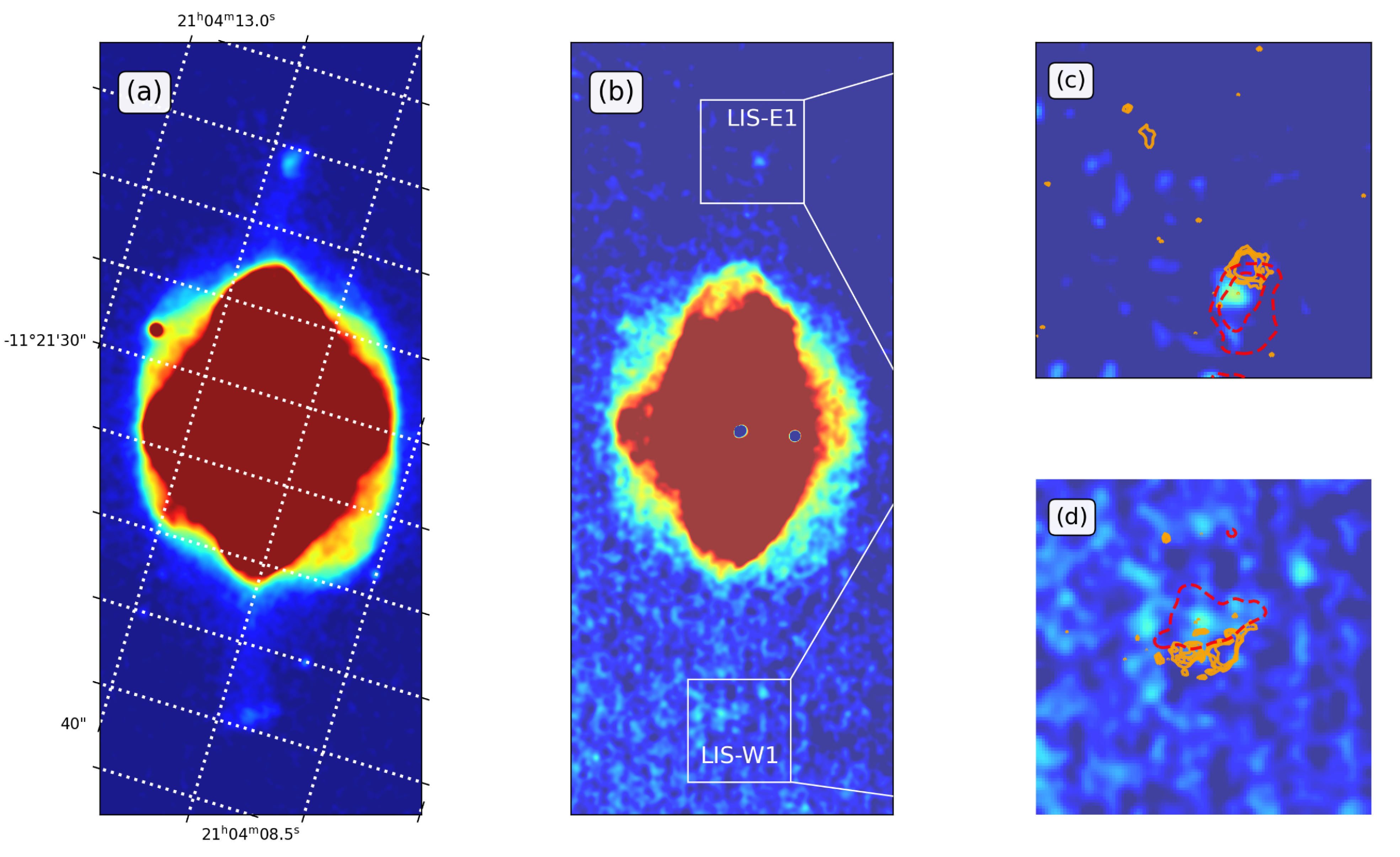}
\caption[]{Gemini NIRI images of NGC~7009. Panel (a): Total \ironii+\hydrogeni~line image. Panel (b): Continuum-subtracted \ironii+\hydrogeni~line image. Panels (c and d): Zoom-in of the continuum-subtracted \ironii+\hydrogeni~image to the LIS-W1 and LIS-E1 overlaid by the H$_2$ 1-0 emission (orange contours) and Br$\gamma$ emission (dashed red contours). Panel (a) is on logarithmic scale and panels (b), (c), and (d) on a linear scale.}
\label{fig1}
\end{figure*}

\begin{table*}
\caption{NGC~7009 \ironii~1.644~$\mu$m+\hydrogeni~12-4~1.640~$\mu$m~and \ironii~1.644~$\mu$m flux and intensity.}
\label{table2}
\centering
\begin{tabular}{lccccccc}
\hline
\multicolumn{1}{c}{Name} & \multicolumn{1}{c}{R.A.} & \multicolumn{1}{c}{Dec.} & \multicolumn{1}{c}{\ironii~1.644~$\mu$m+} & \multicolumn{1}{c}{\ironii~1.644~$\mu$m} &  \multicolumn{1}{c}{Br$\gamma$} & \multicolumn{1}{c}{R(\iron)} & \multicolumn{1}{c}{Area}\\
\multicolumn{1}{c}{} & \multicolumn{1}{c}{(J2000.0)} & \multicolumn{1}{c}{(J2000.0)} & \multicolumn{1}{c}{\hydrogeni~12-4~1.640~$\mu$m} & \multicolumn{1}{c}{} &  \multicolumn{1}{c}{} & \multicolumn{1}{c}{} & \multicolumn{1}{c}{(arcsec/pixel)}\\
\hline
\hline
LIS-E &  21:04:12.41 & -11:21:41.69   &  1.87e-15 (4) & 1.07e-15 & 4.20e-15 & 0.25$\pm$0.06 & 1.164$\times$1.164   \\
         &              &             &  5.87e-5       & 3.36e-5 & 1.32e-4 & & (10x10) \\
LIS-W &  21:04:09.10 & -11:21:53.27   &  5.11e-15 ($<$) & $-$ & 6.09e-15 & $-$ & 2.095$\times$1.396\\
         &              &             &  7.43e-5       & $-$ & 8.86e-5 & & (18x12)   \\
LIS-K & 21:04:10.12  & -11:21:46.98  &  5.81e-14 (9) & 1.27e-14 & 2.39e-13 & 0.05$\pm$0.01 & 3.491$\times$1.745   \\
         &              &             &  4.05e-4       & 8.85e-5 &  1.67e-3 & & (30x15)  \\
\hline
\end{tabular}
\tablefoot{
Fluxes are in erg s$^{-1}$ cm$^{-2}$ (first row) and intensities in erg s$^{-1}$ cm$^{-2}$ sr$^{-1}$ (second row). Numbers in parentheses give the S/N for each flux, while the symbol "$<$" indicates an upper limit.}
\end{table*} 

The f/6 configuration was selected for this narrow-band survey because of the large angular size of the targets providing a field of view of 120$\times$120 arcsec$^2$ and a pixel size of 0.117~arcsec. Several individual frames were obtained for each target with exposure times between 170 and 180~s. The observations were carried out under poor weather conditions, which introduces extra uncertainty on the absolute fluxes. The seeing during the observations varied between 0.6 and 1.0~arcsec. For some PNe, it was necessary to degrade images quality in order to properly subtracted the continuum emission. The observing log is summarised in Table~\ref{table1}.

The thermal emission, dark current and hot pixels on the detector were corrected using the darks and GCAL frames. In order to reduce the total time of the observations, the major axis of each nebula was oriented in the up-down direction on the detector and the dithering was carried out across the left-right direction (minor axis of the nebula). The images were flux-calibrated observing standard stars with the same configuration as the science data. 

Before started the reduction process, the first frames from each sequence were excluded as recommended by Gemini staff. The {\sc python} routines {\sc clearir.py} and {\sc nirlin.py} were applied to all the frames for the correction of the vertical stripping and the non-linearity of the detector. The reduction of the data was then performed with the GEMINI {\sc iraf}\footnote{IRAF is distributed by the National Optical Astronomy Observatories, which are operated by the Association of Universities for Research in Astronomy, Inc., under cooperative agreement with the National Science Foundation.} package for NIRI. The routines Nprepare, Nisky, Niflat, Nireduce, and Imcoadd were used for each imaging set accordingly.

\begin{figure}
\centering
\includegraphics[scale=0.35]{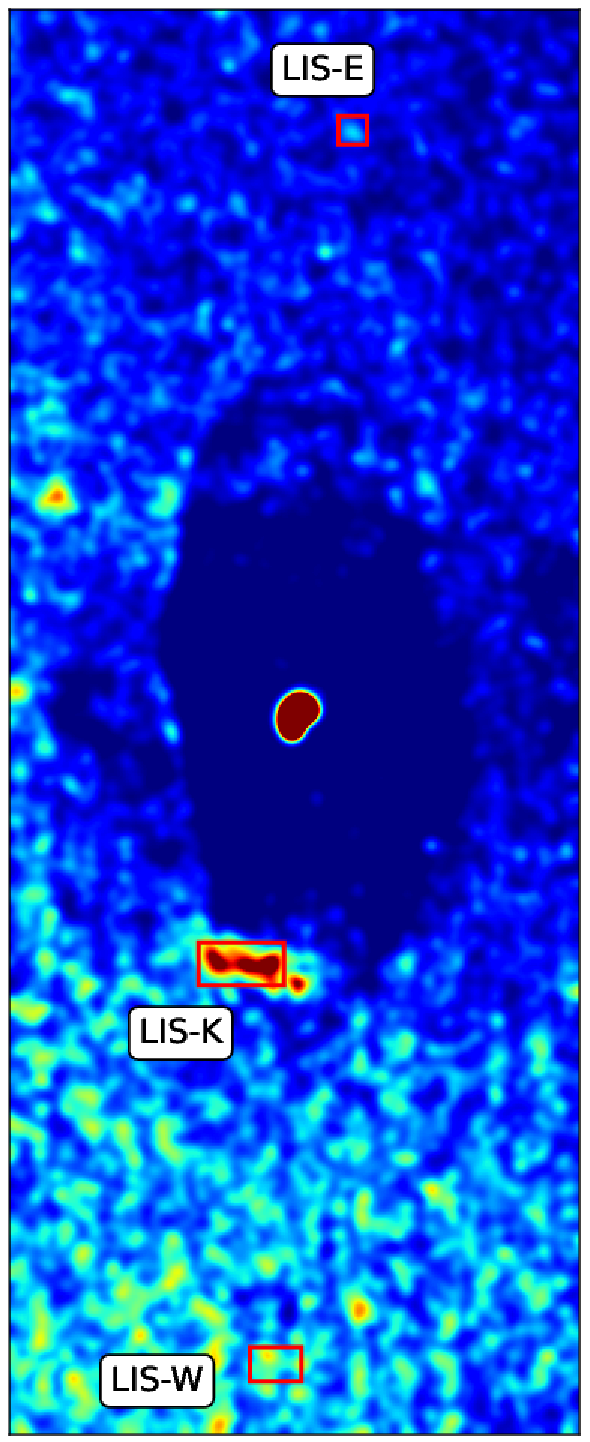}
\caption[]{\ironii~1.644~$\mu$m image of NGC~7009 on a linear intensity scale. Regions that the line fluxes and intensities are measured from are indicated with red rectangles.}
\label{fig2}
\end{figure}

\section{The [Fe~{\sc ii}]~1.644~$\mu$m~narrow-band~imagery}
The first \ironii+\hydrogeni~12-4 line images for a sample of five PNe with NIRI@Gemini are presented with the aim to detect the \ironii~1.644~$\mu$m emission line associated with their LISs and provide further insights into the formation and excitation mechanisms of these intriguing microstructures.

\subsection{NGC~7009}

\begin{figure*}
\vbox{
\includegraphics[scale=0.265]{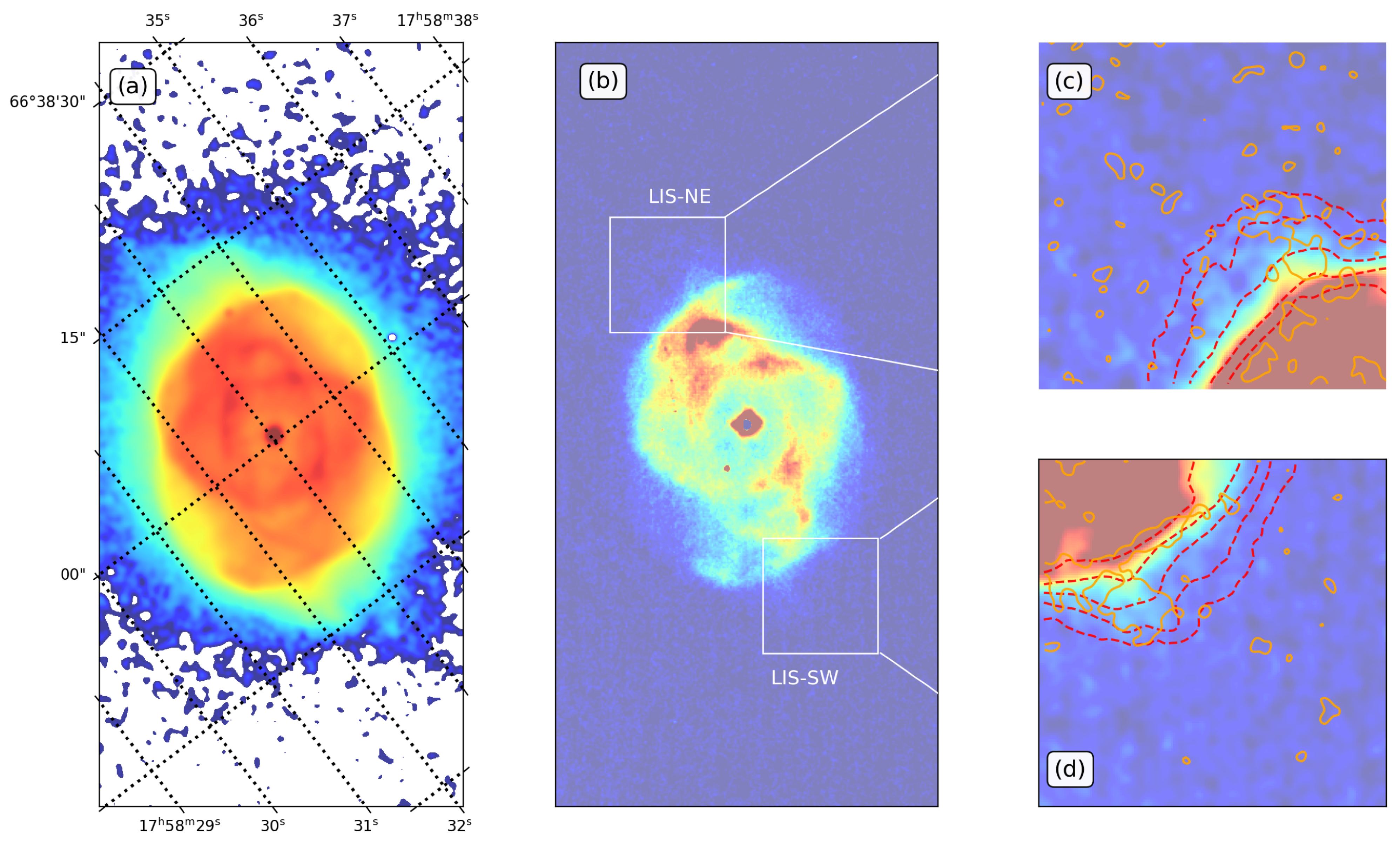}}
\caption[]{Gemini NIRI images of NGC~6543. Panel (a): Br$\gamma$~line image. Panel (b): Continuum-subtracted \ironii+\hydrogeni~line image. Panels (c and d): Zoom-in of the continuum-subtracted \ironii+\hydrogeni~image to the LIS-W1 and LIS-E1 overlaid by the H$_2$ 1-0 emission (black contours) and Br$\gamma$ emission (dashed red contours. Panel (a) is on logarithmic scale and panels (b), (c), and (d) on a linear scale.}
\label{fig3}
\end{figure*}

The strong emission of low ionisation lines such as \nitrogen, \oxygeni, and \sulfurt~ detected at the tips of the jet-like structures in NGC~7009 has been intriguing scientists for years. The recent detection of H$_2$ emission from the same LISs has raised new questions about their nature. The observed H$_2$/Br$\gamma$ and H$_2$ 1-0/2-1 line ratios can be attributed either to UV-pumping or collisional excitation mechanism \citep{Akras2020}. 

The \ironii~1.644~$\mu$m emission line was not detected in the K-band spectra of this nebula published by \cite{Hora1999}. Probably, the slit did not cover the LISs, as stated by the authors. It should be noted that  the H$_2$ 2.12$\mu$m emission line had not been previously detected in the LISs of this nebula either \citep[see ][]{Latter2000}.

In Fig.\ref{fig1}, we present the total and continuum-subtracted \ironii+\hydrogeni~12-4 image of NGC 7009. Strong emission is detected in the main inner nebula and it is accounted for in the \hydrogeni~12-4~1.640~$\mu$m line. Interestingly, a fainter emission is easily perceptible at the eastern LIS but only marginally detected at the western LIS. In both cases, the emission is engulfed by the more extended Br$\gamma$ emission (dashed red contours in panels c and d). This implies that part (if not all) of  the \ironii+\hydrogeni~12-4 emission  may originate from the ionised hydrogen gas. It should be noted the offset of approximately 1~arcsec between the \ironii+\hydrogeni~12-4 and H$_2$ emission (black contours in panels c and d) found for both LISs, with the \ironii+\hydrogeni~12-4 emission laying closer to the central star. The emission flux of the eastern LISs is determined 1.87$\times$10$^{-15}$ erg~s$^{-1}$~cm$^{-2}$, with a signal-to-noise ratio (S/N) of 4 while for the western LIS only an upper limit is presented $<$5.11$\times$10$^{-15}$ erg~s$^{-1}$~cm$^{-2}$.

To calculate how much of the observed \ironii+\hydrogeni~12-4 flux corresponds to the \hydrogeni~12-4 line, the Br$\gamma$ image of NGC~7009 from \cite{Akras2020} was used to construct the theoretical \hydrogeni~12-4 image\footnote{Both \ironii+\hydrogeni~12-4 and Br$\gamma$ images were obtained with the same telescope (Gemini), the same instrument (NIRI) and same configuration. Hence, the subtraction of the images using the central star of the nebula for their alignment becomes straightforward with small error of 2-3 pixels.} (\hydrogeni~12-4/Br$\gamma\sim$0.19) and subtracted it from the observed \ironii+\hydrogeni~12-4 image. 

The first \ironii~1.644~$\mu$m image of NGC~7009 is presented in Fig.~\ref{fig2}, showing a residual emission in the eastern LIS and a barely visible emission in the western LIS. \ironii~1.644~$\mu$m emission is also unveiled in an inner LIS/knot \citep[][]{goncalves2003}. The \ironii~1.644~$\mu$m fluxes are estimated as 1.1 and 12.7 ($\times$10$^{-15}$)~erg~s$^{-1}$~cm$^{-2}$ for the LIS-E and LIS-K, respectively. The intensity varies from 3.36 to 8.85 ($\times$10$^{-5}$) erg~s$^{-1}$~cm$^{-2}$~sr$^{-1}$ (Table~\ref{table2}).

\subsection{NGC~6543}

NGC~6543 is the second PN in our sample for which H$_2$ emission is associated with LISs \citep{Akras2020}. For this PN, the observed H$_2$/Br$\gamma$ and H$_2$ 1-0/2-1 S(1) line ratios indicate a non-thermal origin (UV-pumping mechanism). The \ironii~1.644~$\mu$m emission line was not detected in the K-band spectroscopic data of the southern knot of NGC~6543 \citep{Hora1999}.

The Br$\gamma$ (panel a) and continuum-subtracted \ironii+\hydrogeni~12-4 (panel b) images of NGC~6543 are presented in Fig.~\ref{fig3}. Both emission lines show very similar spatial distribution, clearly displaying the ellipse E105 (major axis at P.A.=105~degrees), the edges of the ellipse E25 (major axis at P.A.=25~degrees), and the \nitrogen~caps \citep[for the definition of these structures see,][]{Reed1999}. Panels (c) and (d) show a zoom-in to the regions of the NE and SW LISs, respectively (see caption in Fig.~\ref{fig3}). 

\begin{figure}
\centering
\includegraphics[scale=0.35]{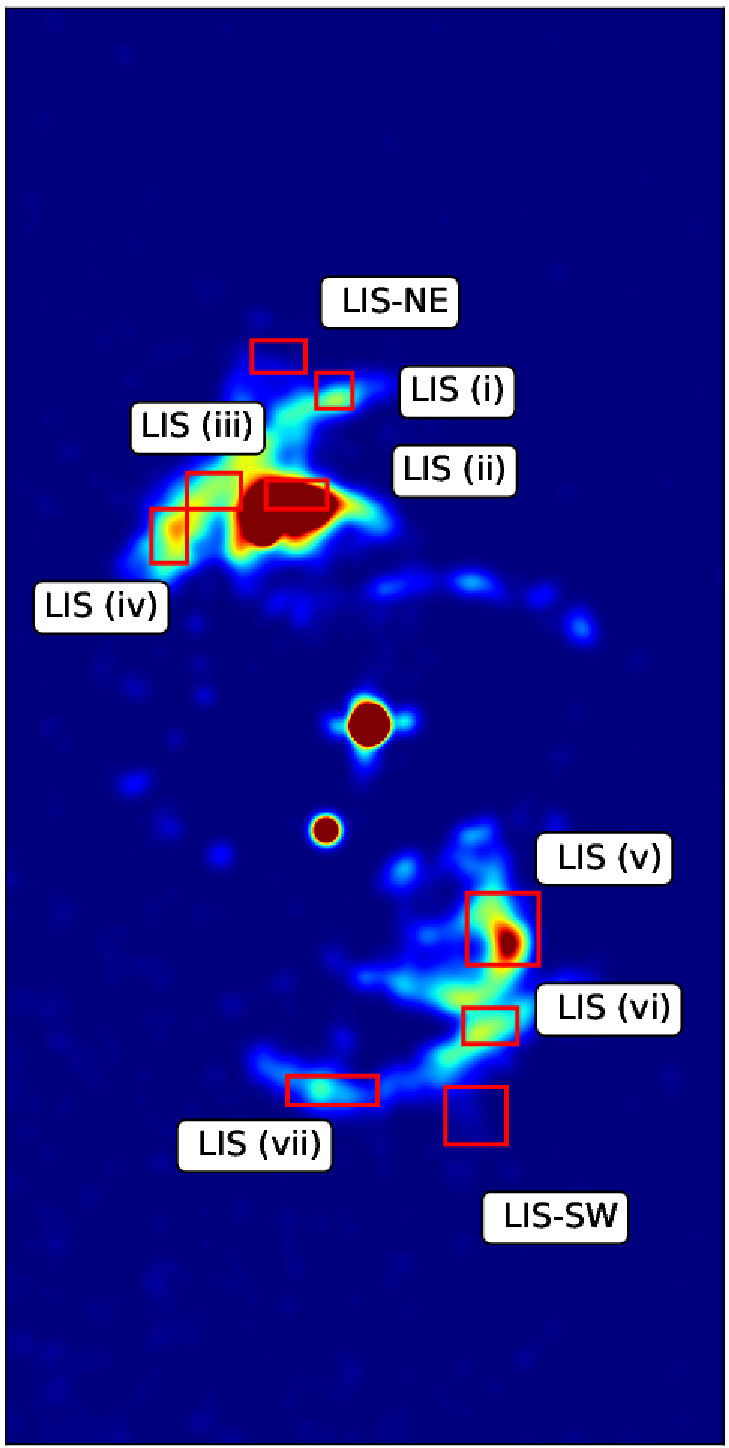}
\caption[]{\ironii~1.644~$\mu$m image of NGC~6543 on a linear intensity scale. Regions that the line fluxes and intensities are measured from are indicated with red rectangles.}
\label{fig4}
\end{figure}

\begin{table*}
\caption{NGC~6543 \ironii~1.644~$\mu$m+\hydrogeni~12-4~1.640~$\mu$m and \ironii~1.644~$\mu$m~fluxes and intensities.}
\label{table3}
\centering
\begin{tabular}{lccccccc}
\hline
\multicolumn{1}{c}{Name} & \multicolumn{1}{c}{R.A.} & \multicolumn{1}{c}{Dec.} & \multicolumn{1}{c}{\ironii~1.644~$\mu$m+} & \multicolumn{1}{c}{\ironii~1.644~$\mu$m} & \multicolumn{1}{c}{Br$\gamma$} & \multicolumn{1}{c}{R(\iron)} &\multicolumn{1}{c}{Area}\\
\multicolumn{1}{c}{} & \multicolumn{1}{c}{(J2000.0)} & \multicolumn{1}{c}{(2000.0)} & \multicolumn{1}{c}{\hydrogeni~12-4~1.640~$\mu$m} & \multicolumn{1}{c}{} &\multicolumn{1}{c}{} &\multicolumn{1}{c}{} & \multicolumn{1}{c}{(arcsec/pixel)}\\
\hline
\hline
LIS-NE &  17:58:33.80 & +66:38:11.72   &  7.86e-15 (5) & 2.82e-15 & 2.65e-14 & 0.11$\pm$0.02 & 1.723$\times$1.002 \\
         &              &                &  1.94e-4       & 6.95e-5 &  6.53e-4 & &  (15x9) \\
LIS-SW  &  17:58:32.85 & +66:37:47.80  &  1.22e-14 ($<$) & $-$ &  5.59e-14 & $-$ & 3.150$\times$2.760  \\
         &              &                &  5.93e-5       & $-$ & 2.73e-4 & & (27x24) \\
LIS (i) &  17:58:33.96 & +66:38:09.80  &  2.02e-14 (14)  & 8.05e-15 & 6.37e-14 & 0.13$\pm$0.01 & 1.164$\times$1.164  \\
         &              &                &  6.32e-4       & 2.52e-4 &  2.01e-3 & & (10x10) \\
LIS (ii) &  17:58:33.44 & +66:38:08.11  &  9.25e-14 (22)  & 4.08e-14 & 2.72e-13 & 0.15$\pm$0.01 &  1.978$\times$0.931  \\
         &              &                &  2.13e-3       & 9.42e-4 &  6.28e-3 & & (17x8) \\
LIS (iii) &  17:58:33.12 & +66:38:09.72  &  5.32e-14 (19)  & 1.85e-14 & 1.82e-13 & 0.10$\pm$0.01 & 1.745$\times$1.164 \\
         &              &                &  1.11e-3       & 3.88e-4 &  3.82e-3 & & (15x10)  \\
LIS (iv) &  17:58:32.76 & +66:38:09.32  &  5.20e-14 (19)  & 2.11e-14 & 1.63-13 & 0.13$\pm$0.01 & 1.164$\times$1.745  \\
         &              &                &  1.09e-3       & 4.41e-4 & 3.31e-3 & & (10x15) \\
LIS (v) &  17:58:32.90 & +66:37:52.78 &  1.19e-13 (24)  & 3.26e-14 & 4.54e-13 & 0.07$\pm$0.01 & 2.327$\times$2.327 \\
         &              &                &  9.33e-4       & 2.56e-4 & 3.56e-3 & & (20x20)   \\
LIS (vi) &  17:58:32.52 & +66:37:50.49 &  3.50e-14 (17)  & 7.01e-15 & 1.47e-13 & 0.05$\pm$0.01 &  1.745$\times$1.164    \\
         &              &                &  7.32e-4       & 1.47e-4 &  3.08e-3 & & (15x10) \\
LIS (vii) &  17:58:31.62 & +66:37:51.90 &  2.67e-14 (12)  & 5.78e-15 & 1.10e-13 &  0.05$\pm$0.01 & 2.909$\times$0.931  \\
         &              &                &  4.19e-4       & 9.08e-5 &  1.73e-3 & & (25x8) \\
\hline
\end{tabular}
\tablefoot{
Fluxes are in erg s$^{-1}$ cm$^{-2}$ (first row) and intensities in erg s$^{-1}$ cm$^{-2}$ sr$^{-1}$ (second row). Numbers in parentheses give the S/N for each flux, while the "$<$" symbol indicates an upper limit.}
\end{table*}

\begin{figure*}
\includegraphics[scale=0.555]{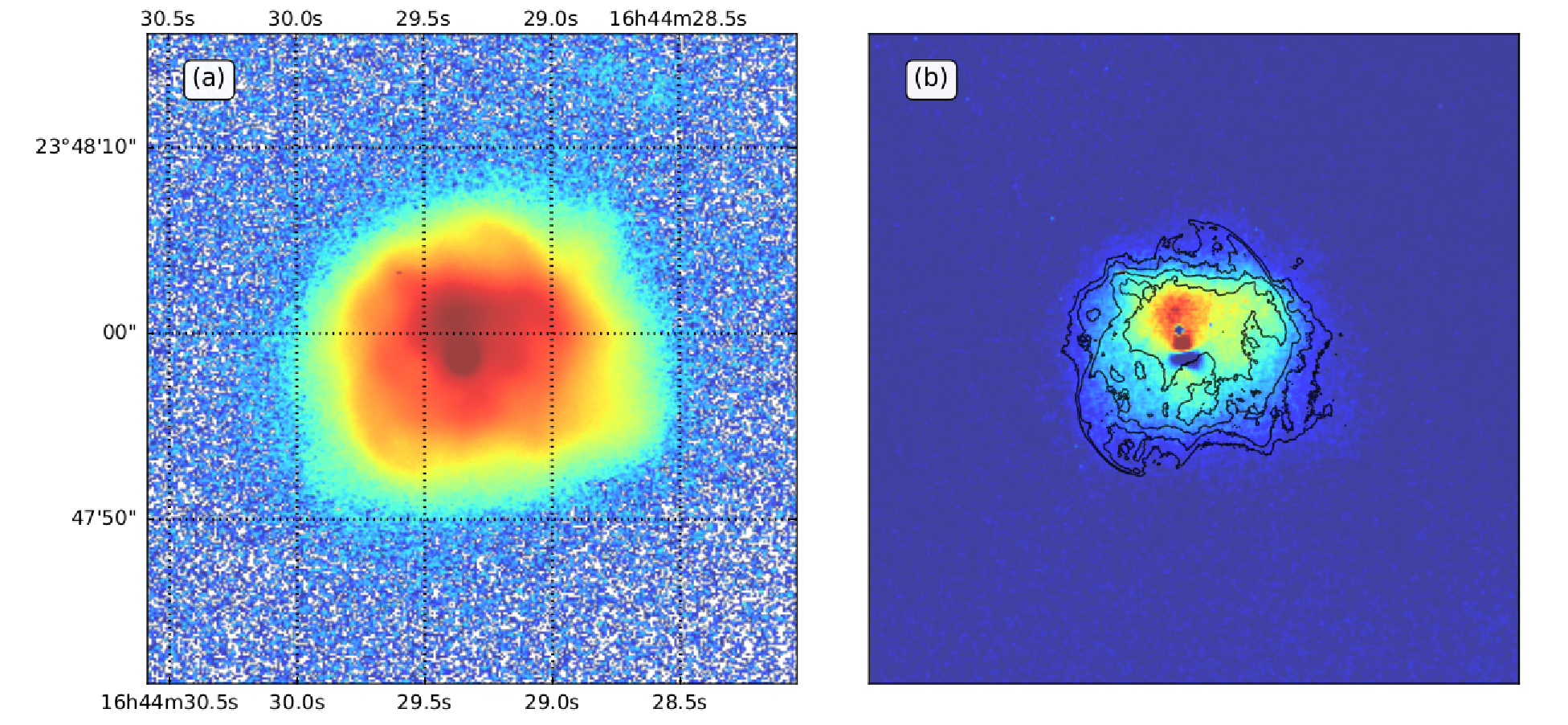}
\caption[]{Gemini NIRI images of NGC~6210. Panel (a): Total \ironii~1.644~$\mu$m+\hydrogeni~12-4~1.640~$\mu$m line image. Panel (b): Continuum-subtracted \ironii~1.644~$\mu$m+\hydrogeni~12-4~1.640~$\mu$m line image overlaid by the HST \nitrogen~6584\AA\ emission (black contours). All panels are on logarithmic scale.}
\label{fig5}
\end{figure*}

\ironii+\hydrogeni~12-4~emission is found to be co-spatial with H$_2$ emission, while Br$\gamma$ is more extended and encloses both emission lines. Fluxes were calculated from the same regions where H$_2$ emission was also measured \citep{Akras2020} and they are equal to 7.86$\times$10$^{-15}$~erg~s$^{-1}$~cm$^{-2}$, with an S/N of 5 for the NE and an upper limit for the SW LIS $<$12.2 $\times$10$^{-15}$~erg~s$^{-1}$~cm$^{-2}$.

\begin{figure*}
\vbox{
\includegraphics[scale=0.475]{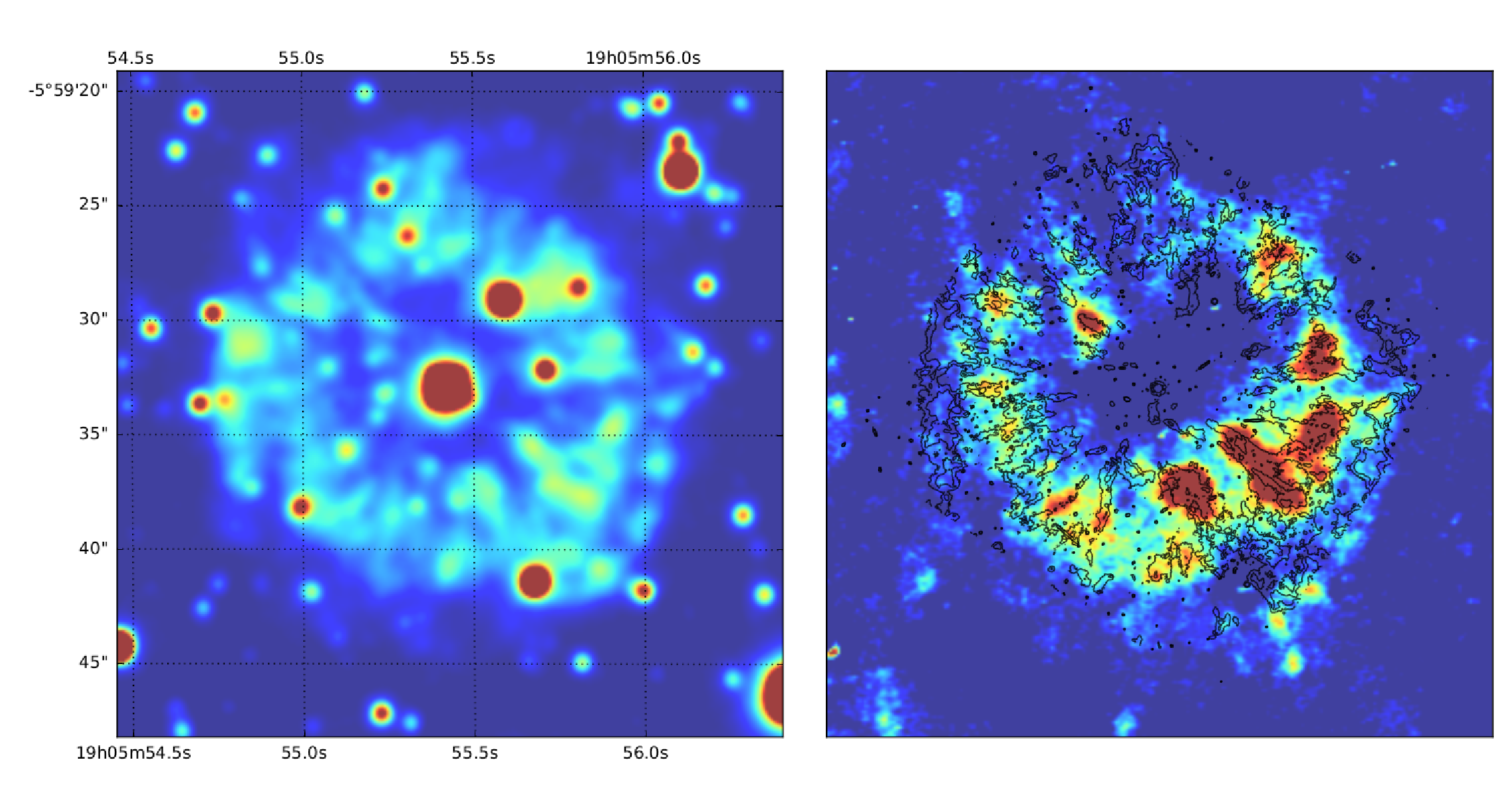}}
\caption[]{Gemini NIRI images of NGC~6751. Panel (a): Total \ironii~1.644~$\mu$m+\hydrogeni~12-4~1.640~$\mu$m line image. Panel (b): Continuum-subtracted \ironii~1.644~$\mu$m+\hydrogeni~12-4~1.640~$\mu$m line image overlaid by the HST \nitrogen~6584\AA\ emission (black contours). All panels are on a linear scale.}
\label{fig6}
\end{figure*}

\begin{figure}
\centering
\includegraphics[scale=0.32]{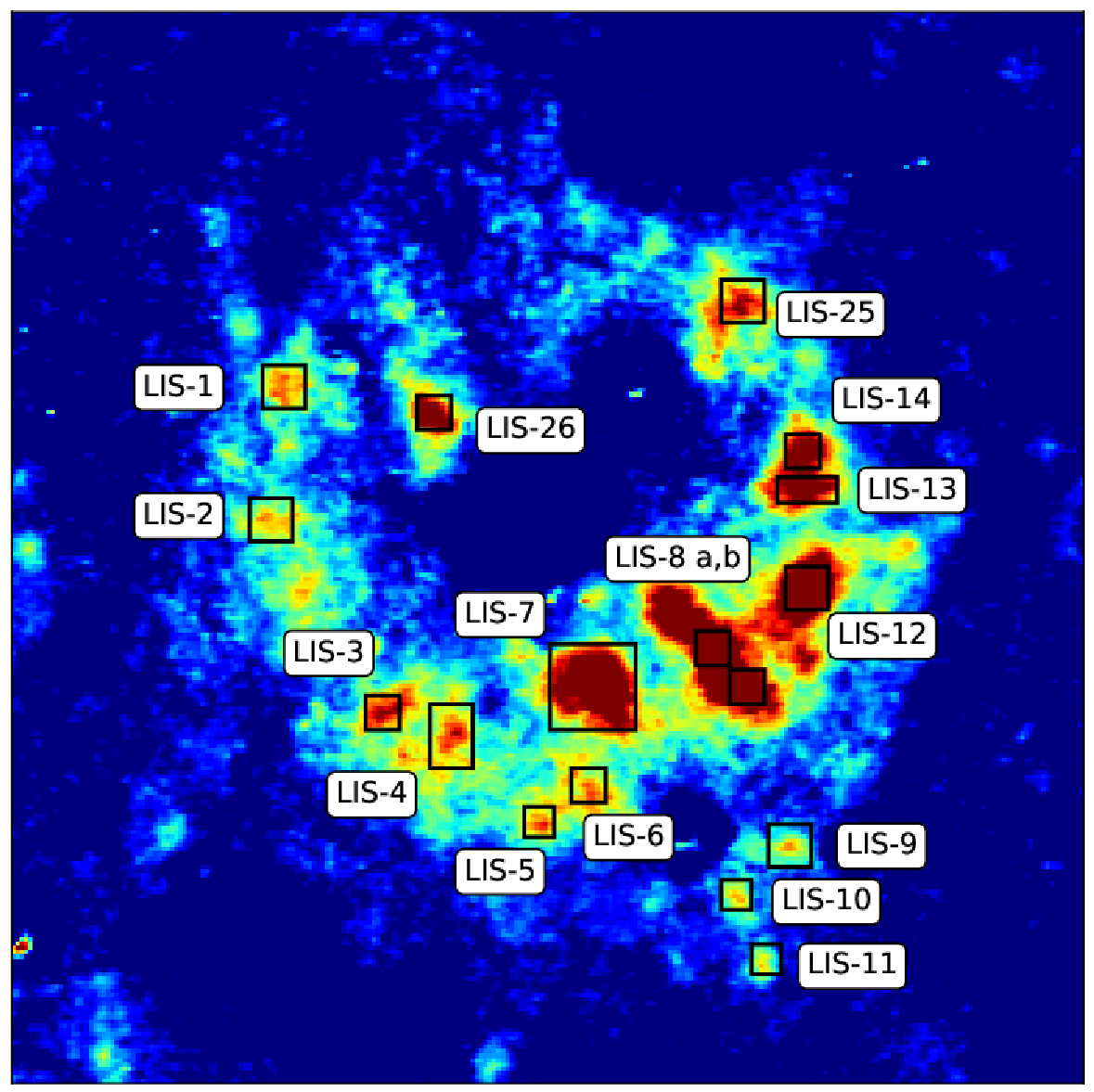}
\caption[]{Continuum-subtracted \ironii~1.644~$\mu$m+\hydrogeni~12-4~1.640~$\mu$m image of NGC~6751 on a linear scale. Regions, from which line fluxes and intensities measured, are indicated with black rectangles.}
\label{fig7}
\end{figure}

Similarly to NGC~7009, the theoretical \hydrogeni~12-4 line image was constructed from the corresponding Br$\gamma$ line image \citep{Akras2020} and subtracted from the total \ironii+\hydrogeni~12-4 image. The resulting \ironii~1.644~$\mu$m image of NGC~6543 is displayed in Fig.~\ref{fig4} with significant emission emanate from the edges of the ellipse E25 and the \nitrogen~caps structures. A weak emission is also detected in the ellipse E105 and the NE/SW LISs. 

\ironii~1.644~$\mu$m line fluxes are measured for various regions in the nebula (see Fig.~\ref{fig4}) ranging from 2.82$\times$10$^{-15}$ to 4.08$\times$10$^{-14}$~erg~s$^{-1}$~cm$^{-2}$ and the intensity from 6.95$\times$10$^{-5}$ to 9.42 $\times$10$^{-4}$~erg~s$^{-1}$~cm$^{-2}$~sr$^{-1}$, respectively (Table~\ref{table3}). 

\subsection{NGC~6210}

NGC~6210, also known as the Turtle-Shape nebula, is morphologically characterised by a bright inner part with several arcs, filaments, and much fainter collimated outflows \citep{Bohigas2015,Rechy2020}.

The total (right panel) and continuum-subtracted (left panel)  \ironii+\hydrogeni~12-4 images of NGC~6210 are shown in Fig.~\ref{fig5}. Emission emanates mainly from the bright inner part of the nebula, without any association with the outflows. The \ironii+\hydrogeni~12-4 line image of NGC~6210 is not flux calibrated because no suitable standard stars were observed. Besides the overall spatial distribution of the \ironii+\hydrogeni~12-4 emission, any link between \ironii~1.644~$\mu$m emission and the LISs distributed in the inner part of the nebula cannot be confirmed or ruled out yet.

\subsection{NGC~6751}
NGC~6751 is a complex multi-shell planetary nebula \citep{Clark2010}. The most intriguing part is the inner shell or ring fragmented into knots and filaments, similar to NGC~2392 nebula \citep{GarciaDiaz2012}.

The total and continuum-subtracted \ironii+\hydrogeni~12-4~images of NGC~6751 are presented in Fig.~\ref{fig6} (panels a and b). The knotty and filamentary structure of the nebula is easily discerned in both images, despite the lower spatial resolution compared to the HST \nitrogen\ image. Panel (b) displays the \ironii+\hydrogeni~12-4~continuum subtracted image with the \nitrogen~$\lambda$6584 contours (black) from the HST image overlaid, showing a very good spatial matching. Most of the knots are detected in the \ironii+\hydrogeni~12-4 and \nitrogen~emission lines. \ironii+\hydrogeni~12-4 flux has been determined for several regions in NGC~6751 varying by an order of magnitude. The faintest and brightest structures have fluxes of 1.26$\times$10$^{-15}$ and 3.94$\times$10$^{-14}$~erg~s$^{-1}$~cm$^{-2}$, respectively (see Table~\ref{table4}).

The absence of a Br$\gamma$ line image prevents us from constructing a theoretical image of \hydrogeni~12-4 line and, as a consequence, the \ironii~1.644~$\mu$m line image of NGC~6751 as well. Nevertheless, we used the integrated dereddened \hb\ flux (2.79$\times$10$^{-13}$~erg~s$^{-1}$~cm$^{-2}$) measured at the C1 region with size of 0.5x10.8~arcsec$^2$ \citep{Chu1991} to obtain the expected flux of the \hydrogeni~12-4 line adopting the Case B recombination theory, n$_{\rm e}$=10$^4$~cm$^{-3}$ and T$_{\rm e}$=10$^4$~K (Br$\gamma$/\hb~$\sim$0.0275, \hydrogeni~12-4~/~Br$\gamma$~$\sim$0.19). Following this approximation, the Br$\gamma$ and \hydrogeni~12-4 fluxes were calculated for each aperture, as well as the expected flux of the \ironii~1.644~$\mu$m line (Table~\ref{table4}). The \ironii~1.644~$\mu$m flux varies from 2$\times$10$^{-15}$ to 3.79$\times$10$^{-14}$~erg~s$^{-1}$~cm$^{-2}$ depending on the size and the position of each region. The regions considered for the \ironii~1.644~$\mu$m flux measurements are depicted in Fig.~\ref{fig7}. These values must be used with caution due to the uncertain estimation of the \hydrogeni~12-4 fluxes. However, there is no doubt about the detection of the \ironii~1.644~$\mu$m line in NGC~6751 given that the C1 region is the brightest structure of the nebula \citep{Chu1991} and the contribution of the \hydrogeni~12-4 line is very likely overestimated. The high R(\iron) values strongly support the presence of shocks in NGC~6751.

\begin{table*}
\caption{NGC~6751 \ironii~1.644~$\mu$m+\hydrogeni~12-4~1.640~$\mu$m and \ironii~1.644~$\mu$m~fluxes and intensities.}
\label{table4}
\centering
\begin{tabular}{lccccccc}
\hline
\multicolumn{1}{c}{Name} & \multicolumn{1}{c}{R.A.} & \multicolumn{1}{c}{Dec.} & \multicolumn{1}{c}{\ironii~1.644~$\mu$m+} & \multicolumn{1}{c}{\ironii~1.644~$\mu$m} & \multicolumn{1}{c}{Br$\gamma$} & \multicolumn{1}{c}{R(\iron)} &\multicolumn{1}{c}{Area}\\
\multicolumn{1}{c}{} & \multicolumn{1}{c}{(J2000.0)} & \multicolumn{1}{c}{(2000.0)} & \multicolumn{1}{c}{\hydrogeni~12-4~1.640~$\mu$m} & \multicolumn{1}{c}{} &\multicolumn{1}{c}{} &\multicolumn{1}{c}{} & \multicolumn{1}{c}{(arcsec/pixels)}\\
\hline
\hline
LIS-1&  19:05:54.95 & -5:59:29.32   & 4.64e-15  (6) & 4.25e-15 & 1.93e-15 & 2.20$\pm$0.37 & 1.164$\times$1.164 \\
         &              &             &  1.46e-4     & 1.33e-4 & 6.06e-5 & & (10x10)  \\
LIS-2&  19:05:54.94 & -5:59:33.02   & 4.52e-15  (6) & 4.13e-15 & 1.93e-15 & 2.14$\pm$0.36 & 1.164$\times$1.164  \\
         &              &             &  1.42e-4     & 1.30e-4 & 6.04e-5 & & (10x10) \\         
LIS-3&  19:05:55.13 & -5:59:38.16   & 5.23e-15  (10) & 4.98e-15 & 1.23e-15 & 4.05$\pm$0.41 & 0.931$\times$0.931   \\
         &              &             &  2.57e-4     & 2.44e-4 & 6.04e-5 & & (8x8)\\ 
LIS-4&  19:05:55.26 & -5:59:38.82   & 1.01e-14  (11) & 9.52e-15 & 2.89e-15 & 3.29$\pm$0.29 & 1.164$\times$1.745  \\
         &              &             &  2.11e-4     & 1.99e-4 & 6.04e-5 & & (10x15)\\ 
LIS-5&  19:05:55.41 & -5:59:41.23   & 3.29e-15  (8) & 3.09e-15 & 9.41e-16 & 3.28$\pm$0.29 & 0.814$\times$0.814   \\
         &              &             &  2.12e-4     & 1.98e-4 & 6.04e-5 & & (7x7)\\ 
LIS-6&  19:05:55.50 & -5:59:40.22   & 4.43e-15  (9) & 4.18e-15 & 1.23e-15 & 3.40$\pm$0.38 & 0.931$\times$0.931 \\
         &              &             &  2.17e-4     & 2.05e-4 & 6.04e-5 & & (8x8)\\ 
LIS-7 &  19:05:55.50 & -5:59:37.45   & 3.94e-14  (13) & 3.79e-14 & 7.69e-15 & 4.92$\pm$0.38 & 2.327$\times$2.327  \\
         &              &             &  3.09e-4     & 2.97e-4 & 6.04e-5 & &(20x20) \\          
LIS-8a&  19:05:55.73 & -5:59:36.37   & 7.85e-15  (12) & 7.59e-15 & 1.23e-15 & 6.17$\pm$0.51 & 0.931$\times$0.931   \\
         &              &             &  3.85e-4     & 3.73e-4 & 6.04e-5 & & (8x8)\\ 
LIS-8b&  19:05:55.80 & -5:59:37.42   & 8.17e-15  (12) & 7.92e-15 & 1.23e-15 & 6.43$\pm$0.54 & 0.931$\times$0.931  \\
         &              &             &  4.01e-4     & 3.88e-4 & 6.04e-5 & & (8x8) \\       
LIS-9&  19:05:55.86 & -5:59:41.81   & 4.54e-15  (6) & 4.15e-15 & 1.93e-15 & 2.15$\pm$0.36 & 1.164$\times$1.164   \\
         &              &             &  1.43e-4     & 1.29e-4 & 6.04e-5 & & (10x10)\\ 
LIS-10&  19:05:55.77 & -5:59:43.14   & 2.63e-15  (7) & 2.44e-15 & 9.41e-16 & 2.60$\pm$0.37 & 0.814$\times$0.814  \\
         &              &             &  1.69e-4     & 1.57e-4 & 6.04e-5 &  &(7x7) \\ 
LIS-11&  19:05:55.82 & -5:59:44.87   & 2.26e-15  (6) & 2.07e-15 & 9.41e-16 & 2.20$\pm$0.37 & 0.814$\times$0.814  \\
         &              &             &  1.45e-4     & 1.33e-4 & 6.04e-5 & & (7x7) \\ 
LIS-12&  19:05:55.89 & -5:59:34.87   & 1.37e-14  (17) & 1.34e-14 & 1.93e-15 & 6.94$\pm$0.41 & 1.164$\times$1.164 \\
         &              &             &  4.33e-4     & 4.21e-4 & 6.04e-5 & & (10x10)  \\ 
LIS-13&  19:05:55.91 & -5:59:32.05   & 8.16e-15  (12) & 7.84e-15 & 1.62e-15 & 4.84$\pm$0.41 & 1.629$\times$0.698   \\
         &              &             &  3.05e-4     & 2.93e-4 & 6.04e-5 & & (14x6)\\  
LIS-14&  19:05:55.89 & -5:59:30.94   & 6.25e-15  (10) & 6.01e-15 & 1.23e-15 & 4.89$\pm$0.49 & 0.931$\times$0.931  \\
         &              &             &  3.07e-4     & 2.95e-4 & 6.04e-5 & &  (8x8)  \\  
LIS-15&  19:05:55.79 & -5:59:26.93   & 7.62e-15  (10) & 7.23e-15 & 1.93e-15 & 3.75$\pm$0.38 & 1.164$\times$1.164  \\
         &              &             &  2.39e-4     & 2.27e-4 & 6.04e-5 & &  (10x10)\\    
LIS-16&  19:05:55.22 & -5:59:29.98   & 6.23e-15  (11) & 5.98e-15 & 1.23e-15 & 4.86$\pm$0.44 & 0.931$\times$0.931  \\
         &              &             &  3.06e-4     & 2.94e-4 & 6.04e-5 & & (8x8)\\ 
\hline
\end{tabular}
\tablefoot{
Fluxes are in erg s$^{-1}$ cm$^{-2}$ (first row) and intensities in erg s$^{-1}$ cm$^{-2}$ sr$^{-1}$ (second row). Numbers in parentheses give the S/N for each flux.}
\end{table*}

\begin{figure*}
\vbox{
\includegraphics[scale=0.485]{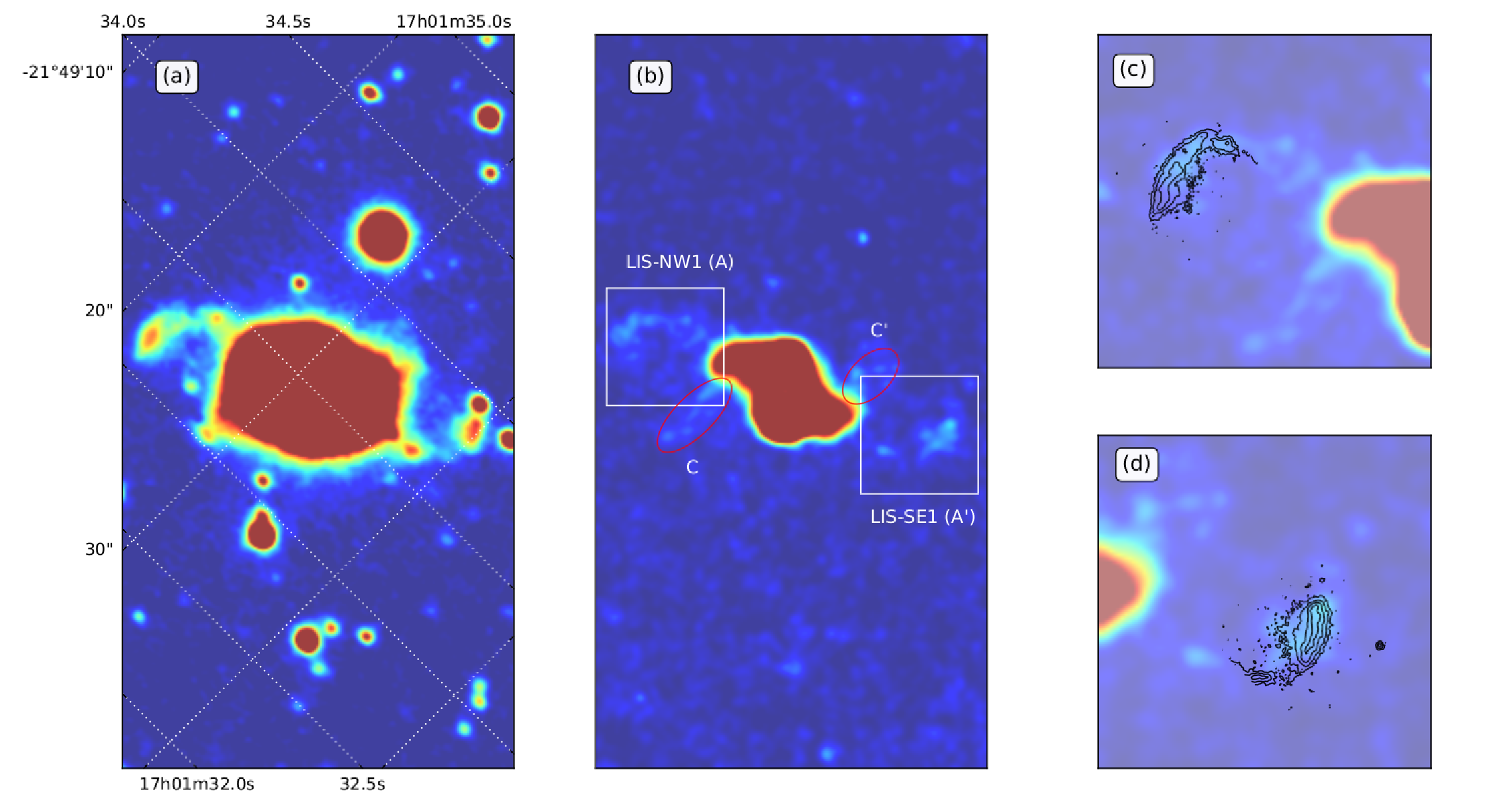}}
\caption[]{Gemini NIRI images of IC~4634. Panel (a): Total \ironii~1.644~$\mu$m+\hydrogeni~12-4~1.640~$\mu$m image. Panel (b): Continuum-subtracted \ironii~1.644~$\mu$m+\hydrogeni~12-4~1.640~$\mu$m image. Panels (c and d): Zoom-in of the continuum-subtracted \ironii~1.644~$\mu$m+\hydrogeni~12-4~1.640~$\mu$m image to the LIS-NW and LIS-SE overlaid by the HST \nitrogen~6584\AA\ emission (black contours). All panels are on a linear scale.}
\label{fig8}
\end{figure*}

\begin{figure}
\centering
\includegraphics[scale=0.220]{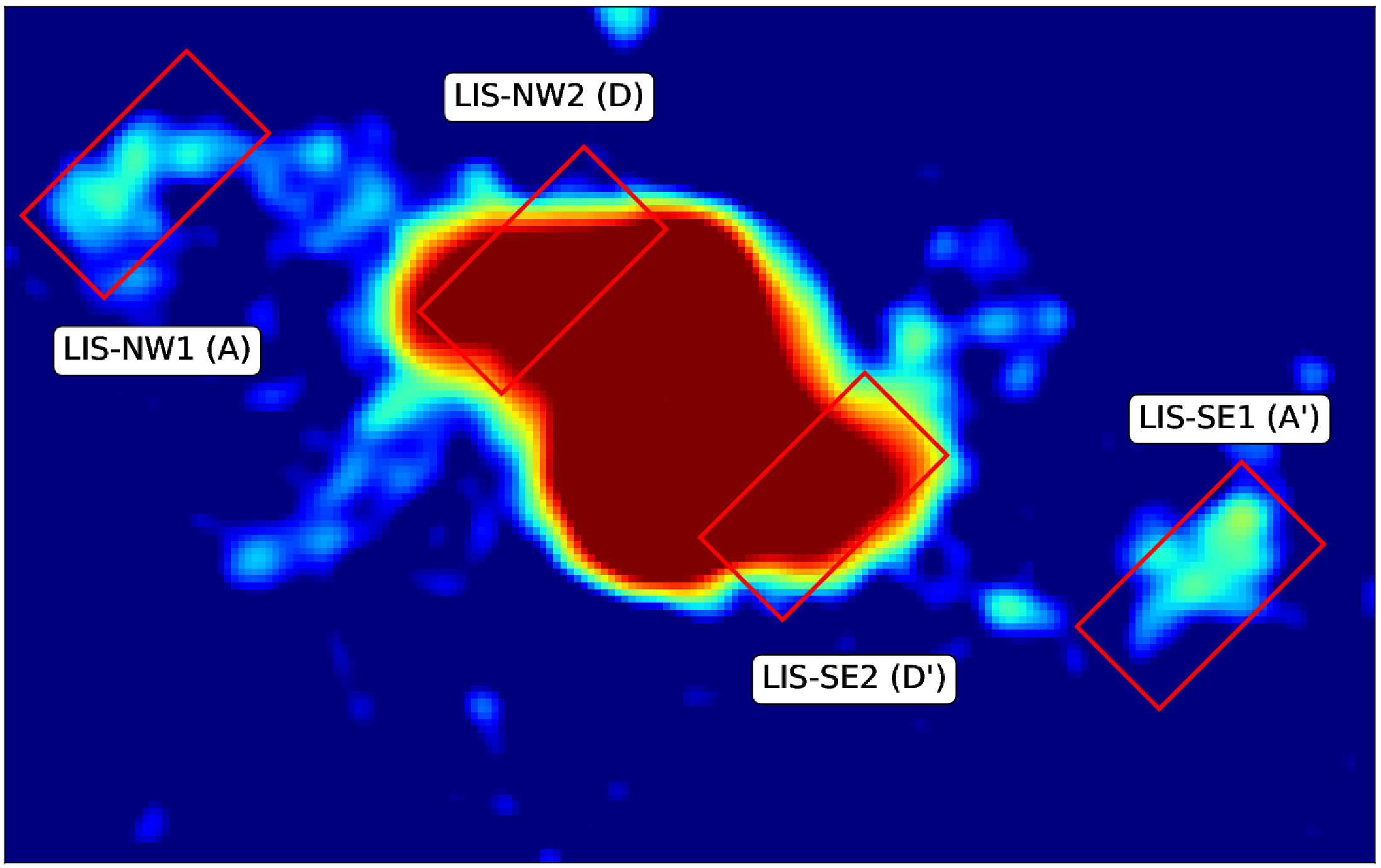}
\caption[]{Continuum-subtracted \ironii~1.644~$\mu$m+\hydrogeni~12-4~1.640~$\mu$m image of IC~4634 on a logarithmic scale. Black rectangular shapes indicate the windows where the line fluxes and intensities are measured.}
\label{fig9}
\end{figure}

\subsection{IC~4634}

The last PN observed in this pilot imaging survey is IC~4634. It is described by an inner and outer S-shape morphology with several individual microstructures bright in low-ionisation lines. Hereafter, we  follow the nomenclature given by \cite{Guerrero2008}: (i) a pair of bow-shock features (A/B and A$^{\prime}$/B$^{\prime}$) forming the outer S-shape, (ii) a pair of bow shock-like features (C/C$^{\prime}$), (iii) a pair of knotty arcs (D/D$^{\prime}$) forming the inner S-shape, and (iv) the bright inner shell.

Figure~\ref{fig8} displays the total and continuum-subtracted \ironii+\hydrogeni~12–4 images of IC~4634 (panels a and b). The S-shape structure of the inner and outer nebula is visible in both images, with a faint \ironii+\hydrogeni~12-4 emission detected in the A/A$^{\prime}$ (white squares). The C/C$^{\prime}$ outflows indicated by red ellipses are barely detected as well. Panels (c) and (d) show a zoom-in view of the A/A$^{\prime}$ features in \ironii+\hydrogeni~12-4~continuum-subtracted image with the HST \nitrogen\ emission line contours overlaid. Both the \ironii+\hydrogeni~12-4 and \nitrogen\ emission lines are co-spatial.

The \ironii+\hydrogeni~12–4 fluxes are estimated for the A/A$^{\prime}$ and D/D$^{\prime}$ features (see Fig.~\ref{fig9}). The D/D$^{\prime}$ features are found to be at least one magnitude brighter (3.79 and 2.69 $\times$10$^{-14}$~erg~s$^{-1}$~cm$^{-2}$) compared to the A/A$^{\prime}$ features (4.30 and 4.87$\times$10$^{-15}$~erg~s$^{-1}$~cm$^{-2}$), with S/N values of 39/28 and 5/6 for each feature, respectively (Table~\ref{table5}). Figure~\ref{fig9} displays the regions (black rectangular shape) where fluxes were calculated.

The absence of Br$\gamma$ imagery or K-band spectroscopy of IC~4634 forced us to use the H$\beta$ image of the nebula obtained with the HST in order to get the theoretical flux of the Br$\gamma$ and \hydrogeni~12-4~1.640~$\mu$m lines. The \hb\ fluxes from the A/A$^{\prime}$ features as defined in Fig.~\ref{fig9} are 7.55 and 6.15$\times$10$^{-14}$~erg s$^{-1}$ cm$^{-2}$ or 1.43 and 1.33$\times$10$^{-13}$~erg s$^{-1}$ cm$^{-2}$, respectively, after applying a correction for the extinction \cite[0.34,][]{Guerrero2008}. These fluxes are two orders of magnitude lower than those reported by \cite{Guerrero2008} while the \hb\ fluxes from the D/D$^{\prime}$ features are comparable.
The theoretical \hydrogeni~12-4 line fluxes for the A/A$^{\prime}$ features are computed 7.52 and 6.92$\times$10$^{-16}$~erg s$^{-1}$ cm$^{-2}$, respectively. According to this analysis, the flux of the \ironii~1.644~$\mu$m line in the A/A$^{\prime}$ features is approximately 3.5 and 4.2$\times$10$^{-15}$~erg s$^{-1}$ cm$^{-2}$, respectively.

\section{Discussion}

The subgroup of fast moving LISs, known as FLIERs \citep[see][]{Balick1993,Balick1998}, are the ideal targets to search for evidence of shocks in PNe. Their high expansion velocities (supersonic velocities $>$40-50 km~s$^{-1}$) and interactions with the surrounding nebular gas or interstellar medium (ISM) can be responsible for the enhancement of the emission from low-ionisation species such as singly ionised nitrogen, sulphur (\nitrogen~$\lambda\lambda$6548,6584, and \sulfurt~$\lambda\lambda$6716,6731) or neutral oxygen and nitrogen (\oxygeni~$\lambda$6300, \nitrogena~$\lambda$5200) relative to the emission from the surrounding nebular gas \citep[e.g.][]{Balick1993,Phillips1998,Dopita1997,goncalves2003,goncalves2004,Raga2008,Akras2016}.

The outward motion of LISs should result in some additional signs such as bow-shocks, which are not frequently observed though. Nevertheless, indirect evidence of shocks is apparent as the augmentation of the electron temperature \citep{Guerrero2013} or the presence of a strong \ironii~1.644$\mu$m line and high R(\ironii) ratio. 

Deep, spatially resolved images of the \ironii~1.644$\mu$m emission line in NGC~7009, NGC~6543, NGC~6751 and IC~4634 have verified the presence of singly ionised iron in LISs and demonstrated that it is cospatial with other low-ionisation emission lines (e.g. \nitrogen~$\lambda$6584). The R(\iron) ratio has been computed for several LISs in the PNe of our list, and it ranges from 0.5 up to 7. The low values are attributed to UV pumping process and the higher values to shocks.

In particular, the eastern LIS of NGC~7009 has R(\iron)=0.25, four times higher than the value measured in the Orion H~II region \citep{Lowe1979}. Despite the fact that we cannot definitively argue for the presence of shocks, the low R(\iron) value entails only low velocity shock waves. Based on the \oxygeniii/\ha\ line ratio maps from HST and MUSE, it has been found that the outer pair of LISs are characterised by an increased ratio at the outer edges \citep[see, ][]{Guerrero2013,Walsh2018,Akras2022}, which may imply a higher electron temperature due to the shock interaction of LISs with the surrounding gas. It should, however, be noted that the photoelectric heating process from dust grains can also be important in high density structures and provide a reasonable explanation for the enhanced \oxygeniii/\ha\ ratio \citep[e.g. ][]{Dopita2000,vanHoof2004}.

The lowest ratio in our sample is measured for NGC~6543, 0.05$<$R(\iron)$<$0.15 and we argue that UV process is more likely responsible for the ionisation of the gas. The same mechanism has also been proposed to explain the low R(H$_2$) and R(Br$\gamma$) ratios found in this nebula \citep{Akras2020}.


On the other hand, significantly high R(\iron) ratios have been measured for NGC~6751 (2$<$R(\iron)$<$7) and IC~4634 (R(\iron)$\sim$1; A/A$\prime$~features), providing strong evidence that shocks are the dominant mechanism. It is worth noticing that the A/A$^{\prime}$ features in IC~4634 have noticeable higher T$_{\rm e}$(\oxygeniii) by 1000-1500~K, compared to the rest of the components \citep{Hajian1997}, while T$_{\rm e}$(\nitrogen) is the same in all nebular structures. This temperature difference may be associated with the interaction of the A/A$^{\prime}$ features with the ISM. The high velocities up to 70-100~\kms~measured at the A/A$^{\prime}$ (features 1 and 5 in \citealt{Hajian1997}) support this hypothesis. From a statistical point of view, \citealt{Belen2023b} have also shown that T$_{\rm e}$(\oxygeniii) is systematically higher than T$_{\rm e}$(\nitrogen) in LISs.


\begin{table*}
\caption{IC~4634 \ironii~1.644~$\mu$m+\hydrogeni~12-4~1.640~$\mu$m and \ironii~1.644~$\mu$m fluxes and intensities.}
\label{table5}
\centering
\begin{tabular}{lccccccc}
\hline
\multicolumn{1}{c}{Name} & \multicolumn{1}{c}{R.A.} & \multicolumn{1}{c}{Dec.} & \multicolumn{1}{c}{\ironii~1.644~$\mu$m+} & \multicolumn{1}{c}{\ironii~1.644~$\mu$m} &  \multicolumn{1}{c}{Br$\gamma$} & \multicolumn{1}{c}{R(\iron)} & \multicolumn{1}{c}{Area}\\
\multicolumn{1}{c}{} & \multicolumn{1}{c}{(J2000.0)} & \multicolumn{1}{c}{(J2000.0)} & \multicolumn{1}{c}{\hydrogeni~12-4~1.640~$\mu$m} & \multicolumn{1}{c}{} &  \multicolumn{1}{c}{} & \multicolumn{1}{c}{} & \multicolumn{1}{c}{(arcsec)}\\
\hline
\hline
LIS-NW2 (A) & 17:01:33.22 & -21:49:22.09  &  4.30e-15 (5) & 3.57e-15 & 3.96e-15 & 0.91$\pm$0.18 & 1.978$\times$3.955  \\
         &              &                &  2.34e-5       &1.95e-5 & 2.15e-5 & & (17x34) \\
LIS-NW2 (D) & 17:01:33.50 & -21:49:28.47 &  3.79e-14 (39) & 2.24e-14 & 8.18e-14 & 0.27$\pm$0.01 & 1.978$\times$3.955 \\
         &              &                &  2.06e-4       & 1.22e-4 & 4.45e-4 & &  (17x34) \\
LIS-SE2 (D$^{\prime}$) & 17:01:33.56 & -21:49:34.56 &  2.69e-14 (28) & 1.68e-14 & 5.29e-14 & 0.32$\pm$0.01 & 1.978$\times$3.955 \\
         &              &                &  1.46e-4       & 9.22e-5 &  2.88e-4 & &  (17x34)  \\
LIS-SE1 (A$^{\prime}$) & 17:01:33.83 & -21:49:39.71  &  4.87e-15 (6) & 4.18e-15 & 3.64e-15 & 1.15$\pm$0.19 & 1.978$\times$3.955  \\
         &              &                &  2.65e-5      & 2.28e-5 &  1.98e-5 & & (17x34)\\
\hline
\end{tabular}
\tablefoot{
Fluxes are in erg s$^{-1}$ cm$^{-2}$ (first row) and intensities in erg s$^{-1}$ cm$^{-2}$ sr$^{-1}$ (second row). Numbers in parentheses give the S/N for each flux, while the "$<$" symbol indicates an upper limit.}
\end{table*}

Overall, four out of five PNe (NGC~7009, NGC~6543, NGC~6571, and IC~4634) observed in this pilot \ironii+\hydrogeni~12-4 imaging survey provide sufficient evidence to claim for the first detection of the \ironii~1.644~$\mu$m line in the LISs of these PNe. However, the absolute fluxes of the \ironii~1.644~$\mu$m line computed in this work are characterised by high uncertainties due to the accumulative uncertainties in all the steps followed to correct the observed emission for the contribution of the \hydrogeni~12-4 line (H$\beta$ and Br$\gamma$ fluxes, n$_{\rm e}$=10$^4$~cm$^{-3}$, and T$_{\rm e}$=10$^4$~K assumptions for case B, the different weather conditions between the observations, the different instruments, slit positions, and extracted windows).

Scrutinising the available NIR spectroscopic data from PNe by \cite{Hora1999,Lumsden2001} and correcting for the contribution of the \hydrogeni~12-4~1.640~$\mu$m line, we find that the R(\iron) ratio ranges from 0.06 up to 4. Five out of 15 PNe or 33~percent have R(\iron)$<$0.1, which is indicative of photo-heated gas. Shocks can only be attributed to a handful of PNe or specific nebular structures with R(\iron)$>$1 (e.g. the south-east region in Hb~12, the northern knot in M~2-9, the north-east region in NGC~2440, CRL~618, M~1-78, and BD~+30~3639). It should be noted though that the aforementioned PNe exhibit R(H$_2$)$<$4, which is usually attributed to the UV-pumping process. There are also two PNe that have R(H$_2$)$>$8 and may be associated with shock, whereas their low R(\iron) is only twice large the Orion's value ($\sim$0.13). Photoionisation and shock models produce a wide range of R(H$_2$) with significant overlapping that  impedes the differentiation between two mechanisms \citep[see Fig. 6 in][ and references therein]{Akras2020}.

Our detection of the \ironii~1.644~$\mu$m line depends on the  estimate and subtraction of the \hydrogeni~12-4~1.640~$\mu$m. To corroborate our detection, we searched the literature for the potential detection of other optical Fe lines. Several lines, from a singly ionised and up to six times ionised Fe (either collisionally excited or recombined) have been reported for the four PNe; for instance, \ironii~$\lambda$5261, $\lambda$7154, $\lambda$8617, and/or \ironiii~$\lambda$4658, $\lambda$4701, $\lambda$4881, $\lambda$5270, $\lambda$5412\footnote{Due to the presence of the \helium~$\lambda$5412, this detection is highly uncertain.}, among others
\citep[NGC~7009; ][]{Fang2018}, \citep[NGC~6543; ][]{Perinotto1999,Hyung2000}, \citep[IC~4634; ][]{Hyung1999,Guerrero2008}, and \citep[NGC~6751; ][]{Chu1991}.

We note that these identifications are not all directly associated with the LISs, where the near-IR \ironii~1.644$\mu$m line is detected, except for the line $\lambda$5270 from the A$^{\prime}$ feature in IC~4634 and the line $\lambda$8617 from the north \nitrogen~cap feature in NGC~6543. We have to mention that the \ironii~$\lambda$8617 line has also been detected in the MUSE data of NGC~7009 and specifically in the eastern LIS. The theoretical \ironii~1.644~$\mu$m/8617~\AA~line ratio is $\sim$9, 5 and 1 for N$_e$= 10$^3$, 10$^4$ and 10$^5$~cm$^{-3}$, respectively \citep{Koo2016}. The \ironii~$\lambda$8617 line flux from the MUSE data is computed $\sim$1.5e$^{-16}$~erg~cm$^{-2}$~s$^{-1}$, which yields a ratio of $\sim$7 (Bouvis et al. in prep.). This ratio implies an electron density close to 10$^3$~cm$^{-3}$, which agrees with the observations.


The comparison of the \ironii~1.644~$\mu$m intensities with the predictions from the shock models of \cite{Hollenbach1989} has pointed out to the following model, a high density (10$^4$ cm$^{-3}$) shock model with velocities between 80 and 100~\kms~can explain the average intensity of the \ironii~1.644~$\mu$m line ($\sim$2.10$\times$10$^{-4}$~erg s$^{-1}$ cm$^{-2}$ sr$^{-1}$) of NGC~6751, as well as the average intensity of the \nitrogen~caps in NGC~6543 ($\sim$3.60$\times$10$^{-4}$~erg s$^{-1}$ cm$^{-2}$ sr$^{-1}$). Regarding the eastern LIS of NGC~7009, we conclude that two shock models (i) n$\sim$10$^4$~cm$^{-3}$ and v$\sim$50~\kms, or (ii) n$\sim$10$^3$~cm$^{-3}$ and v$\sim$80~\kms~can both explain the observed intensity of the \ironii~1.644~$\mu$m line ($\sim$3.3$\times$10$^{-5}$~erg s$^{-1}$ cm$^{-2}$ sr$^{-1}$). Despite the fact that \cite{Hollenbach1989} did not provide the predictions for a denser molecular gas, we presume that a slower shock of $\sim$30~\kms\ is likely able to reproduce the observed intensities of the \ironii\ line in denser gas. The observed intensities of the H$_2$ 2.12$\mu$m line from the eastern LIS in NGC~7009 \citep{Akras2020} are higher than the two models predicted, but comparable with the prediction from a highly dense gas ($\sim$10$^5$~cm$^{-3}$). High densities between 10$^3$ and 10$^5$~cm$^{-3}$ have been reported for the shell and knots of NGC~7009 by \cite{Hyung2023}. Regarding the H$\beta$ emission from all the aforementioned structures, shock models with velocity $<$100~\kms~and densities between 10$^{3}$ to 10$^{7}$~cm$^{-3}$ can reproduce the observed intensity \cite[see Table~1 in][]{Hollenbach1989}. It is important to point out that slow shock models require  denser gas. Based on the more recent shock models by \cite{Koo2016}, the observed \ironii~1.644~$\mu$m line fluxes can be reproduced either by fast velocity shock models (90-100~km~s$^{-1}$) and low pre-shock density of 10~cm$^{-3}$, or shock models with velocities lower than 80~km~s$^{-1}$ and higher pre-shock density of $\sim$100~cm$^{-3}$. All these models are characterised by R(\ironii)$>$1, which increases for higher velocities. 

Thus, we suggest that the shock interaction of the LISs with the low density surrounding gas or ISM requires high velocities, which are not always observed. On the contrary, a slow shock wave has to interact with  denser gas in order to have the same emission. In the first case, the interaction of LISs with the surrounding nebular due to their outwards motion, the H$_2$ 2.12$\mu$m line should lie at a larger distance behind the shock front than the \ironii\ 1.644$\mu$m line \citep[see Fig.~1 in ][]{Hollenbach1989} or emanate from at least the same region behind the shock front \citep{Novikov2018,Novikov2019}.  However, our observations show the exact opposite result, along with an offset between the two lines of $\sim$1~arcsec. 

An alternative scenario to explain the enhanced emission from the optical low-ionisation lines, \ironii~1.644~$\mu$m, and H$_2$~2.12$\mu$m is the photoevaporation process of clumps illuminated by a strong UV radiation field. This process has been discussed by several authors \citep[e.g.][]{Sternberg1989,Burton1990,Storzer2000}. For instance, \cite{Mellema1998} performed a detailed analysis of photoevaporated clumps in PNe carrying out both an analytic and a numerical approach. This process offers an adequate explanation for several of the LISs' characteristics \citep[see also ][]{Raga2005}. According to the PDR models by \cite{Burton1990}, a high dense clump (10$^{6-7}$~cm$^{-3}$) illuminated by a strong UV radiation field (G$_0$=10$^{4-5}$)\footnote{G$_0$ gives the intensity of the UV radiation field in terms of the Habing field, and it is equal to 1.6$\times$10$^{-3}$ erg~cm$^{-2}$~s$^{-1}$ \citep{Habing1968}.} can reproduce the intensity of the H$_2$ lines as well as the high line ratio because of collisional de-excitation, measured at the eastern LIS in NGC~7009 \citep{Akras2020}. Photoevaporation process has also been proposed to explain the cometary shapes of knots in the Helix nebula \citep[e.g.][]{LopexMartin2001}.

More specifically, the photons from the central star (or UV source) interact with the inner face of the clump and start evaporating the molecular gas. This evaporating flow moves backward and interacts with the stellar wind or the nebular gas. As a result, a reverse shock wave is formed moving in the opposite direction and propagates through the clump. The pressure that drives the shock wave depends on the ionising photon flux of the central star. According to the photoevaporation model of \cite{Mellema1998}, the velocity of the reverse shock can be as high as 20-30~\kms\ for a central source of 50000~K and 7000~L$\odot$. Such a slow velocity shock interacts with the highly dense molecular code of the LISs (i.e. in the second scenario) and this is likely to be the mechanism that leads to the observed emission spectrum. Moreover, the same shock wave can also lead to the destruction of the dust grains and the release of Fe to gas-phase which would explain our detections. Meanwhile, the high density of molecular core of the LISs provide the necessary shield from being fully dissociated. The hypothesis of the photoevaporation can also account for the observed offset between the H$_2$ and \ironii~lines, with the latter being closer to the central star.

Two noteworthy outcomes of the photoevaporation process are (i) the atomic hydrogen emission (Br$\gamma$) is more extended than the \nitrogen\ emission (or other line) due to the ionisation of the evaporated flow \citep[see fig.7][]{Mellema1998} and (ii) the spatial offset between the Br$\gamma$ and H$_2$ emission lines. Such a spatial offset between the two  hydrogen lines has been observed in molecular clouds \citep{Hartigan2015,Carlsten2018} as well as in the outer pair of LISs in NGC~7009 \citep{Akras2020}.

\section{Conclusions}

The \ironii~narrow-band imaging survey of five PNe with LISs was carried with NIRI@Gemini and presented here. Three emission lines, \ironii~1.644~$\mu$m, \hydrogeni~12-4 1.640~$\mu$m, and \silicon~1.645~$\mu$m, were covered by the Gemini filter. \silicon~1.645~$\mu$m is in general very faint in PNe and its contribution is negligible, while \hydrogeni~12-4 1.640~$\mu$m is relatively strong (0.5~percent of \hb); thus, it was taken into account in order to obtain the \ironii~1.644~$\mu$m fluxes.

The \ironii~1.644~$\mu$m + \hydrogeni~12-4 1.640~$\mu$m blended emission was detected in all five PNe. The contribution of the \hydrogeni~12-4 1.640~$\mu$m line was determined from the available Br$\gamma$ images or \hb~fluxes. After correcting for the contribution of the \hydrogeni~12-4 1.640~$\mu$m line, the detection of \ironii~1.644~$\mu$m line was confirmed in four out of five PNe. In all cases, it is found to be directly associated with LISs.

Beside the detection of the \ironii~1.644~$\mu$m line, its direct association with shocks is not confirmed. Based on the \ironii~1.644~$\mu$/Br$\gamma$ line ratio, we argue that the \ironii~1.644~$\mu$m emission is ascribed to photoionisation process in NGC~6543 ($<$0.15) and to shocks in IC~4634 ($\sim$1) and NGC~6571 (between 2 and 7), attributed to the high ratios. Both mechanisms are seen in the case of the eastern LIS in NGC~7009 ($<0.25$).

The observed displacement between the \ironii, H$_2$, and Br$\gamma$ lines is an indicative of the photoevaporation process in the LISs, while it contradicts the predictions from shock interaction with the nebular gas or ISM. It is likely that a slow-moving shock in the LISs is responsible for the release of Fe in gas phase and the observed emission lines. The presence of such a shock wave ought to be validated with an observational confirmation.

\begin{acknowledgements}
We would like to thank the anonymous referee for the constructive comments that helped  us to improve the quality of the paper. The authors would like to thank Dr. Martin Guerrero for providing us the flux calibrated HST \hb\ image of IC~4634. The research project is implemented in the framework of H.F.R.I call \lq\lq~Basic research financing (Horizontal support of all Sciences)\rq\rq~under the National Recovery and Resilience Plan \lq\lq Greece 2.0\rq\rq~funded by the European Union – NextGenerationEU (H.F.R.I. Project Number: 15665). SA and KB acknowledge support from H.F.R.I. IA acknowledges support from CAPES, Ministry of Education, Brazil, through a PNPD fellowship. DGR acknowledges support from the CNPq grants 428330/2018-5 and 313016/2020-8. GR-L acknowledges support from CONACyT (grant 263373). Based on observations obtained at the international Gemini Observatory, a program of NSF’s OIR Lab [processed using the Gemini IRAF package and DRAGONS (Data Reduction for Astronomy from Gemini Observatory North and South)], which is managed by the Association of Universities for Research in Astronomy (AURA) under a cooperative agreement with the National Science Foundation. on behalf of the Gemini Observatory partnership: the National Science Foundation (United States), National Research Council (Canada), Agencia Nacional de Investigaci\'{o}n y Desarrollo (Chile), Ministerio de Ciencia, Tecnolog\'{i}a e Innovaci\'{o}n (Argentina), Minist\'{e}rio da Ci\^{e}ncia, Tecnologia, Inova\c{c}\~{o}es e Comunica\c{c}\~{o}es (Brazil), and Korea Astronomy and Space Science Institute (Republic of Korea). The following software packages in Python were used: Matplotlib \citep{Hunter2007}, NumPy \citep{Walt2011}, SciPy \citep{SciPy2020} and AstroPy Python \citep{Astropy2013,Astropy2018}.\end{acknowledgements}

\section*{Data availability}
The data presented in this work will be available at the CDS or upon request to the corresponding author. The observations are also available in the Gemini Science Archive.

%
%
\bibliographystyle{aa}
\bibliography{references}

\end{document}